
\input harvmac.tex


\input epsf.tex
\def\figin{\epsfcheck\figin}\def\figins{\epsfcheck\figins}
\def\epsfcheck{\ifx\epsfbox\UnDeFiSIeD
\message{(NO epsf.tex, FIGURES WILL BE IGNORED)}
\gdef\figin##1{\vskip2in}\gdef\figins##1{\hskip.5in}
\else\message{(FIGURES WILL BE INCLUDED)}%
\gdef\figin##1{##1}\gdef\figins##1{##1}\fi}
\def\DefWarn#1{}
\def\figinsert{\goodbreak\midinsert}
\def\ifig#1#2#3{\DefWarn#1\xdef#1{fig.~\the\figno}
\writedef{#1\leftbracket fig.\noexpand~\the\figno}%
\figinsert\figin{\centerline{#3}}\medskip\centerline{\vbox{\baselineskip12pt
\advance\hsize by -1truein\noindent\footnotefont{\bf Fig.~\the\figno:} #2}}
\bigskip\endinsert\global\advance\figno by1}

\lref\greg{   G.~W.~Moore, M.~R.~Plesser and S.~Ramgoolam,
  ``Exact S matrix for 2-D string theory,''
  Nucl.\ Phys.\ B {\bf 377}, 143 (1992)
  [arXiv:hep-th/9111035].
}

\def\half{ { 1\over 2}}
\lref\piljin{
  T.~Suyama and P.~Yi,
  ``A holographic view on matrix model of black hole,''
  JHEP {\bf 0402}, 017 (2004)
  [arXiv:hep-th/0401078].
}

\lref\dasgupta{
  S.~Dasgupta and T.~Dasgupta,
  ``Renormalization group approach to c = 1 matrix model on a circle and
  D-brane decay,''
  arXiv:hep-th/0310106.
}

\lref\minahanpol{
  J.~A.~Minahan and A.~P.~Polychronakos,
  ``Interacting fermion systems from two-dimensional QCD,''
  Phys.\ Lett.\ B {\bf 326}, 288 (1994)
  [arXiv:hep-th/9309044].
   J.~A.~Minahan and A.~P.~Polychronakos,
  ``Integrable systems for particles with internal degrees of freedom,''
  Phys.\ Lett.\ B {\bf 302}, 265 (1993)
  [arXiv:hep-th/9206046].
  J.~A.~Minahan and A.~P.~Polychronakos,
  ``Density correlation functions in Calogero-Sutherland models,''
  Phys.\ Rev.\ B {\bf 50}, 4236 (1994)
  [arXiv:hep-th/9404192].
}

\lref\polreview{
  A.~P.~Polychronakos,
  ``Generalized statistics in one dimension,''
  arXiv:hep-th/9902157.
}

\lref\kos{
  I.~K.~Kostov,
  ``String equation for string theory on a circle,''
  Nucl.\ Phys.\ B {\bf 624}, 146 (2002)
  [arXiv:hep-th/0107247].
  S.~Y.~Alexandrov, V.~A.~Kazakov and I.~K.~Kostov,
  ``Time-dependent backgrounds of 2D string theory,''
  Nucl.\ Phys.\ B {\bf 640}, 119 (2002)
  [arXiv:hep-th/0205079].
   I.~K.~Kostov,
  ``Integrable flows in c = 1 string theory,''
  J.\ Phys.\ A {\bf 36}, 3153 (2003)
  [Annales Henri Poincare {\bf 4}, S825 (2003)]
  [arXiv:hep-th/0208034].
 M.~Aganagic, R.~Dijkgraaf, A.~Klemm, M.~Marino and C.~Vafa,
  ``Topological strings and integrable hierarchies,''
  arXiv:hep-th/0312085.
}

\lref\douglascft{
 M.~R.~Douglas,
  ``Conformal field theory techniques in large N Yang-Mills theory,''
  arXiv:hep-th/9311130.
}

\lref\berkooz{
  M.~Berkooz and M.~Rozali,
  ``Near Hagedorn dynamics of NS fivebranes, or a new universality class  of
  coiled strings,''
  JHEP {\bf 0005}, 040 (2000)
  [arXiv:hep-th/0005047].
}

\lref\thooft{    G.~'t Hooft,
  ``A Two-Dimensional Model For Mesons,''
  Nucl.\ Phys.\ B {\bf 75}, 461 (1974).
}

\lref\seibergliouville{N. Seiberg, notes on liouville}

\lref\gkthermal{  D.~J.~Gross and I.~R.~Klebanov,
  ``One-Dimensional String Theory On A Circle,''
  Nucl.\ Phys.\ B {\bf 344}, 475 (1990).
}

\lref\gkvortex{
  D.~J.~Gross and I.~R.~Klebanov,
  ``Vortices And The Nonsinglet Sector Of The C = 1 Matrix Model,''
  Nucl.\ Phys.\ B {\bf 354}, 459 (1991).
}

\lref\maloog{  J.~M.~Maldacena and H.~Ooguri,
  ``Strings in AdS(3) and SL(2,R) WZW model. I,''
  J.\ Math.\ Phys.\  {\bf 42}, 2929 (2001)
  [arXiv:hep-th/0001053].
}

\lref\fzzbdy{  V.~Fateev, A.~B.~Zamolodchikov and A.~B.~Zamolodchikov,
``Boundary Liouville field theory. I: Boundary state and boundary  two-point
function,''
arXiv:hep-th/0001012.
}

\lref\teschnerbdy{
J.~Teschner,
``Remarks on Liouville theory with boundary,''
arXiv:hep-th/0009138.
}

\lref\mo{ G.~Marchesini and E.~Onofri,
``Planar Limit For SU(N) Symmetric Quantum Dynamical Systems,''
J.\ Math.\ Phys.\  {\bf 21}, 1103 (1980).
}

\lref\fukuda{
 T.~Fukuda and K.~Hosomichi,
``Super Liouville theory with boundary,''
Nucl.\ Phys.\ B {\bf 635}, 215 (2002)
[arXiv:hep-th/0202032].
}

\lref\dkns{
 D.~Kutasov and N.~Seiberg,
  ``Noncritical Superstrings,''
  Phys.\ Lett.\ B {\bf 251}, 67 (1990).
}

\lref\twodbh{ I.~Bars and D.~Nemeschansky,
  ``String Propagation In Backgrounds With Curved Space-Time,''
  Nucl.\ Phys.\ B {\bf 348}, 89 (1991).  I.~Bars,
USC-91-HEP-B3.
  S.~Elitzur, A.~Forge and E.~Rabinovici,
  ``Some global aspects of string compactifications,''
  Nucl.\ Phys.\ B {\bf 359}, 581 (1991).
   K.~Bardacki, M.~J.~Crescimanno and E.~Rabinovici,
  ``Parafermions From Coset Models,''
  Nucl.\ Phys.\ B {\bf 344}, 344 (1990).
    M.~Rocek, K.~Schoutens and A.~Sevrin,
  ``Off-shell WZW models in extended superspace,''
  Phys.\ Lett.\ B {\bf 265}, 303 (1991).
    G.~Mandal, A.~M.~Sengupta and S.~R.~Wadia,
  ``Classical solutions of two-dimensional string theory,''
  Mod.\ Phys.\ Lett.\ A {\bf 6}, 1685 (1991).
  E.~Witten,
  ``On string theory and black holes,''
  Phys.\ Rev.\ D {\bf 44}, 314 (1991).
}

\lref\kkk{
V.~Kazakov, I.~K.~Kostov and D.~Kutasov,
``A matrix model for the two-dimensional black hole,''
Nucl.\ Phys.\ B {\bf 622}, 141 (2002)
[arXiv:hep-th/0101011].
}

\lref\chatone{
 M.~R.~Douglas, I.~R.~Klebanov, D.~Kutasov, J.~Maldacena, E.~Martinec and N.~Seiberg,
  ``A new hat for the c = 1 matrix model,''
  arXiv:hep-th/0307195.
}
\lref\tato{
T.~Takayanagi and N.~Toumbas,
  ``A matrix model dual of type 0B string theory in two dimensions,''
  JHEP {\bf 0307}, 064 (2003)
  [arXiv:hep-th/0307083].
}

\lref\moore{  G.~W.~Moore,
  ``Gravitational phase transitions and the Sine-Gordon model,''
  arXiv:hep-th/9203061.
}

\lref\berkbh{
 N.~Berkovits, S.~Gukov and B.~C.~Vallilo,
  ``Superstrings in 2D backgrounds with R-R flux and new extremal black
  holes,''
  Nucl.\ Phys.\ B {\bf 614}, 195 (2001)
  [arXiv:hep-th/0107140].
}

\lref\ari{
 A.~Giveon, A.~Konechny, A.~Pakman and A.~Sever,
  ``Type 0 strings in a 2-d black hole,''
  JHEP {\bf 0310}, 025 (2003)
  [arXiv:hep-th/0309056].
  }
\lref\hk{
K.~Hori and A.~Kapustin,
 ``Duality of the fermionic 2d black hole and N = 2 Liouville theory as  mirror
symmetry,''
JHEP {\bf 0108}, 045 (2001)
[arXiv:hep-th/0104202].
}

\lref\kutasovtherm{
  D.~Kutasov and D.~A.~Sahakyan,
  ``Comments on the thermodynamics of little string theory,''
  JHEP {\bf 0102}, 021 (2001)
  [arXiv:hep-th/0012258].
}

\lref\boulkaz{
D.~Boulatov and V.~Kazakov,
``One-dimensional string theory with vortices as the upside down matrix
oscillator,''
Int.\ J.\ Mod.\ Phys.\ A {\bf 8}, 809 (1993)
[arXiv:hep-th/0012228].
}

\lref\fzzdual{L. Fateev, A. Zamolodchikov and Al. Zamolodchikov, unpublished notes.}

\lref\refsomeone{
 I. Klebanov,
 ``String theory in two dimensions",
arXiv:hep-th/9108019;
P. Ginsparg and G. Moore,
 ``Lectures on 2D gravity and 2D string theory",
arXiv:hep-th/9304011;
 J.~Polchinski,
  ``What is string theory?,''
  arXiv:hep-th/9411028.
   }

\lref\zz{
A.~B.~Zamolodchikov and A.~B.~Zamolodchikov,
``Structure constants and conformal bootstrap in Liouville field theory,''
Nucl.\ Phys.\ B {\bf 477}, 577 (1996)
[arXiv:hep-th/9506136].
}

\lref\jmns{ J. Maldacena and N. Seiberg, in preparation.}


\Title{\vbox{\baselineskip12pt \hbox{hep-th/0503112}}}{\vbox{
\centerline {Long strings in two dimensional string theory  }
 \centerline{  and non-singlets in the matrix model}}}

\bigskip
\centerline{ Juan Maldacena }
\bigskip
\centerline{ \it Institute for Advanced Study, Princeton, NJ 08540, USA }

\bigskip

\vskip .3in

We consider two dimensional string backgrounds. We discuss the physics of
long strings that come from infinity. These are related to non-singlets in the dual
matrix model description.

\Date{}
\vfill

\eject

\newsec{ Introduction}

The main motivation for this paper is to get some insights about the
 Lorentzian physics of
the two dimensional black hole \twodbh . A proposal for the matrix model
dual of the  Euclidean black hole was made
 by Kazakov, Kostov and Kutasov  in \kkk . These authors proposed a matrix model description
which involves adding Wilson lines for the ordinary gauged matrix model that
describes the two dimensional string background \refs{\refsomeone}.
This implies that we need to understand the matrix model in its non-singlet sector.
In this paper we study
aspects of the physics of the matrix model in its non-singlet sector.
While we will not give a picture for the two dimensional black hole, the
remarks in this paper might be helpful in this quest. Further work on this correspondence
includes \ref\MooreGA{
  G.~W.~Moore,
  ``Gravitational phase transitions and the Sine-Gordon model,''
  arXiv:hep-th/9203061.
}\piljin \ref\KazakovPJ{
  V.~A.~Kazakov and A.~A.~Tseytlin,
  ``On free energy of 2-d black hole in bosonic string theory,''
  JHEP {\bf 0106}, 021 (2001)
  [arXiv:hep-th/0104138].
}\ref\DavisXB{
  J.~L.~Davis, L.~A.~Pando Zayas and D.~Vaman,
  ``On black hole thermodynamics of 2-D type 0A,''
  JHEP {\bf 0403}, 007 (2004)
  [arXiv:hep-th/0402152].
}\ref\AlexandrovUT{
  S.~Alexandrov,
  ``Matrix quantum mechanics and two-dimensional string theory in  non-trivial
  arXiv:hep-th/0311273.
}\ref\AlexandrovPZ{
  S.~Y.~Alexandrov and V.~A.~Kazakov,
  ``Thermodynamics of 2D string theory,''
  JHEP {\bf 0301}, 078 (2003)
  [arXiv:hep-th/0210251].
}\ref\AlexandrovFH{
  S.~Y.~Alexandrov, V.~A.~Kazakov and I.~K.~Kostov,
  ``Time-dependent backgrounds of 2D string theory,''
  Nucl.\ Phys.\ B {\bf 640}, 119 (2002)
  [arXiv:hep-th/0205079].
}\ref\DavisXI{
  J.~L.~Davis and R.~McNees,
  ``Boundary counterterms and the thermodynamics of 2-D black holes,''
  arXiv:hep-th/0411121.
}\ref\ParkYC{
  J.~Park and T.~Suyama,
  ``Type 0A matrix model of black hole, integrability and holography,''
  arXiv:hep-th/0411006.
}\ref\SenYV{
  A.~Sen,
  ``Symmetries, conserved charges and (black) holes in two dimensional string
  theory,''
  JHEP {\bf 0412}, 053 (2004)
  [arXiv:hep-th/0408064].
}\ari \ref\AlexandrovUH{
  S.~Alexandrov,
  ``Backgrounds of 2D string theory from matrix model,''
  arXiv:hep-th/0303190.
}\ref\AlexandrovCM{
  S.~Alexandrov and V.~Kazakov,
  ``Correlators in 2D string theory with vortex condensation,''
  Nucl.\ Phys.\ B {\bf 610}, 77 (2001)
  [arXiv:hep-th/0104094].
}\ref\ulf{U.~H.~Danielsson, J.~P.~Gregory, M.~E.~Olsson, P.~Rajan and M.~Vonk,
  ``Type 0A 2D black hole thermodynamics and the deformed matrix model,''
  JHEP {\bf 0404}, 065 (2004)
  [arXiv:hep-th/0402192].
  }.

We will show that the double scaling limit of the matrix model in
its simplest non-trivial representation, the adjoint
representation, is related to a two dimensional string background
with a folded string that comes in from the weak coupling region
and goes back to the weak coupling region. Such a string is not
static. Indeed,  we get a time dependent configuration. In the
double scaling limit, the system does not have a well defined
ground state but it has interesting time-dependent solutions. An
interesting observable is the scattering amplitude for a folded
string that comes from the weak coupling region and goes back to
the weak coupling region\foot{
Folded strings have also been previously studied in \ref\barsone{
  W.~A.~Bardeen, I.~Bars, A.~J.~Hanson and R.~D.~Peccei,
  ``A Study Of The Longitudinal Kink Modes Of The String,''
  Phys.\ Rev.\ D {\bf 13}, 2364 (1976).
   W.~A.~Bardeen, I.~Bars, A.~J.~Hanson and R.~D.~Peccei,
  ``Quantum Poincare Covariance Of The D = 2 String,''
  Phys.\ Rev.\ D {\bf 14}, 2193 (1976).
  } \ref\barsothers{
  I.~Bars,
  ``Folded strings,''
  arXiv:hep-th/9412044.
  I.~Bars,
  ``Folded strings in curved space-time,''
  arXiv:hep-th/9411078.
  I.~Bars and J.~Schulze,
  ``Folded strings falling into a black hole,''
  Phys.\ Rev.\ D {\bf 51}, 1854 (1995)
  [arXiv:hep-th/9405156].
}.}. This phase can be computed
exactly at tree level in string theory using formulas in \fzzbdy .
 We compare this to the matrix model computation. We reduce the
matrix model computation to a rather simple looking eigenvalue
problem. Though we were not able to solve completely this
eigenvalue problem we could show that in the two asymptotic limits
of high and low energies, the answer matches with the
corresponding exact expressions in string theory. Previous work on
the non-singlet sector includes \refs{\gkvortex,\boulkaz}. This
picture of the non-singlet sector is important for understanding
the physical meaning of the results of computations done via
T-duality or the Euclidean worldsheet theory.

We consider the same problem also in the type 0A/0B superstring cases,  where a new feature
arises. There are two answers that depend on the treatment of
 a particular fermion zero mode.

This paper is organized as follows. In section 2 we discuss the physics of long
strings in two dimensional string theory. In section 3 we consider non-singlets
in the hermitian matrix model. We first concentrate on the adjoint representation and
then we discuss some features for general representations. In section 4 we consider
non-singlets in the complex matrix model that is dual to 0A superstring theory.
We end with a discussion.

\newsec{ Long strings in two dimensional string theory}

\subsec{Classical strings in two dimensions}

Let us first consider a simple problem \barsone . Suppose we have a
classical string on a two dimensional spacetime parametrized by $
X^0 , \phi $.  We choose a gauge with $\partial_+ X^0 =
\partial_- X^0=1$ with  $ \partial_\pm = \partial_\tau \pm \partial_\sigma$,
 where
$\tau, \sigma$ are the  worldsheet coordinates.
The classical equations then reduce to $\partial_+ \partial_- \phi =0$ and the
Virasoro constraints
 \eqn\vircon{
 -1 +  (\partial_+ \phi)^2 = 0 ~,~~~~~~~~~-1 + (\partial_- \phi)^2 =0
}
Then we conclude that $\partial_+ \phi =\pm 1$. The first
derivative does not have to be continuous. So the most general
solution has the form $\phi = \phi_+(\sigma^+) + \phi_-(
\sigma^-) $ where $\phi_+$ and $\phi_-$ are periodic functions
with the same period and with derivative piece-wise equal to $\pm
1$. A simple example would be
\eqn\solclp{
\phi_+(s) \equiv \phi_-(-s) \equiv
| s- L/2| ~,~~~~~{\rm  for}~~~ 0 < s <
L
}
and defined in a periodic fashion outside this interval.
This describes a pulsating string. At $X^0=0$ the string is folded and  stretched to
its maximum length $L$ from $\phi=0$ to $\phi=L$. At later times the two folds
  start moving towards each other at the speed of light, they cross at
$X^0=L/2$ and they end end up with a configuration similar to the original one
at $X^0=L$. See figure 1.
We can think of the whole configuration as two massless particles
joined by a string. In fact, the energy levels that we get are similar to those
we get for mesons in 2d QCD with massless quarks at high excitations levels  \thooft
\foot{In 2d QCD we have an open string rather than a folded string but the form
of the solutions is essentially identical. }.
Here we looked
at a simple solution, but the string can have many folds, see \barsone \barsothers\ for
further discussion.

\ifig\constractingstring{ Two dimensional pulsating string. The ends of the string move
at the speed of light.  }
{\epsfxsize 3.0 in\epsfbox{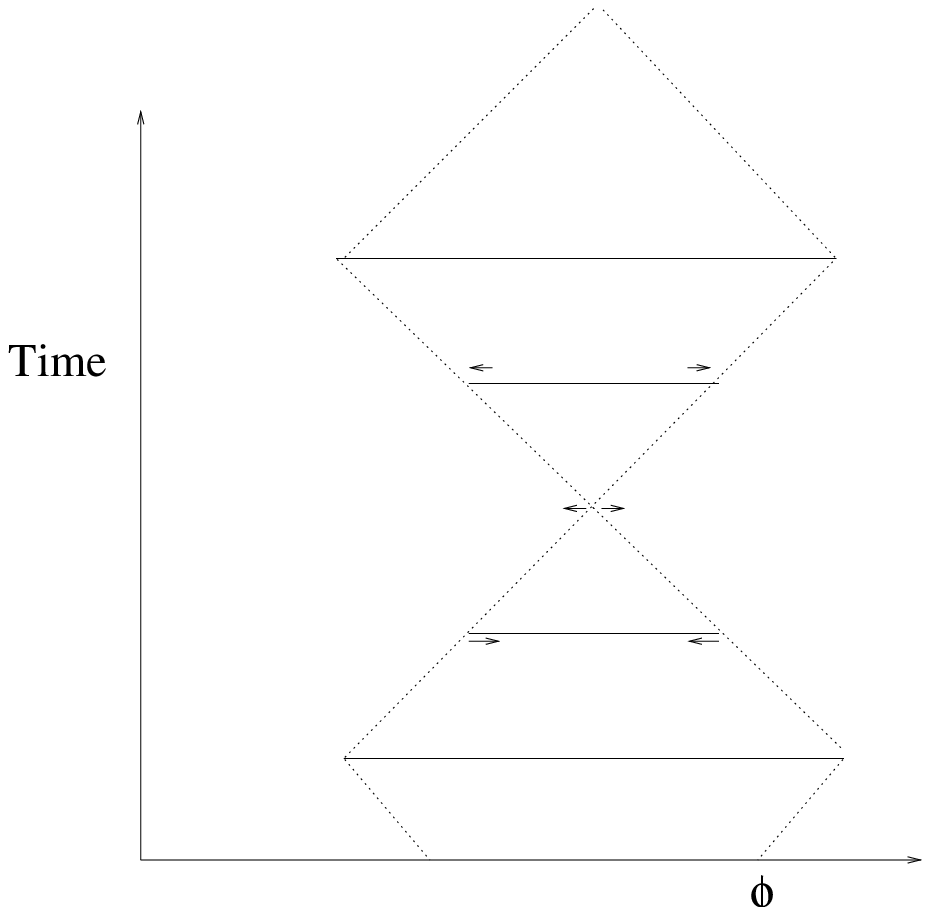}}

\subsec{Long strings with a linear dilaton}

Now let us consider the situation with linear dilaton, $\Phi = Q \phi $.
 Then the
Virasoro constrains in  \vircon\ are replaced by\foot{
Throughout this paper we set $\alpha' =1$ for the bosonic string and
$\alpha' = \half$ for the superstring.}
 \eqn\vircond{
  -1 +  (\partial_+ \phi)^2  - Q\partial_+^2 \phi= 0 ~,~~~~~~~~~
 -1 + (\partial_- \phi)^2 - Q \partial_-^2 \phi =0
}
Solving these equations we find
\eqn\solequ{ \phi = \phi_0 - Q\log \cosh {\sigma^+ \over  Q} - Q \log
\cosh{ \sigma^- \over Q }= \phi'_0 - Q \log ( \cosh { \tau \over Q} + \cosh{
\sigma \over Q} ) }
This
is the most general solution after we allow shifts of $\tau$ and
$\sigma$. The solution corresponding to the usual pointlike string,
where $\phi = \pm \tau$, can be viewed as a limit of this solution.
Except for this special limit,
the solution \solequ\ is not periodic. It  represents a folded string stretched to
minus infinity in the $\phi$ direction. The tip of the
string moves from the weakly coupled region into strong coupling
and back into the weakly coupled region.
One minor difference with the strings discussed in the previous section
 is that in this case  the tip of the
string looks a bit more like a massive particle,
with a mass of order $Q M_s $. After a  few string times
the tip of the string
is accelerated to nearly the speed of light so that its mass becomes
unimportant.
Note that we do not
have oscillating strings. This agrees with the fact that, for $Q=2$,
 there are no
closed string states in the linear dilaton region other than the massless ``tachyon".

\ifig\longdilaton{ Long string coming in from $\phi = -\infty$ and bouncing back. The
vertical direction is time and we show snapshots at different times. The strings stretch
from $\phi = - \infty$, or a large negative cutoff value $\phi_c$. The tip comes from
the left and, after it reaches the
largest value of $\phi =\phi_m$, it starts moving back to the left.  }
{\epsfxsize 2.5 in\epsfbox{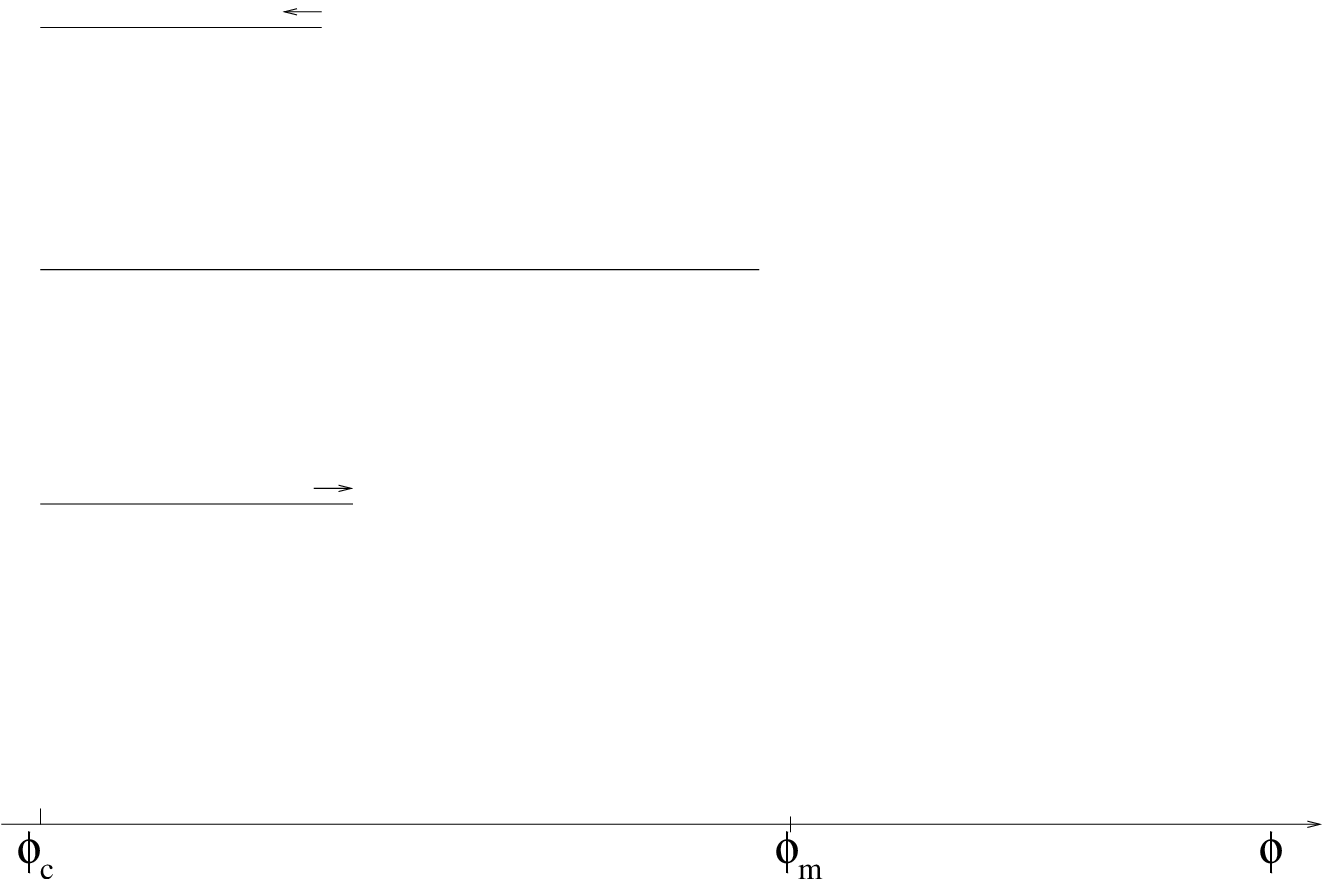}}

We can think of this string as a particle (the tip of the string) with
a linear potential (provided by the string tension).
This enables us to use a WKB approximation for the wavefunction. The particle
momentum and wavefunction are given by
\eqn\classwkb{
 p = \pm 2 { 1 \over2 \pi } (\phi_m - \phi) ~,~~~~~~~~~ \psi \sim e^{ i \int d\phi p} \sim
 e^{ \pm i { 1 \over 2 \pi} (\phi - \phi_m)^2}
 }
where we used the relativistic dispersion relation that is valid for $p \gg 1$. In
\classwkb\ we also used that the force is given by twice the tension since the string
is folded.   Here
 $\phi_m$ is just an integration constant which is the value of $\phi$ at which the tip of the
 string bounces back. This is then related to the energy of the string through
 \eqn\energ{
 E = 2 { 1 \over 2\pi} ( \phi_m - \phi_c ) ~,~~~~~~~~~
  \epsilon \equiv E - E_{div} = { 1 \over \pi} \phi_m ~,~~~~~~~~~~
 E_{div} = - { 1 \over 2 \pi} \phi_c
}
where $\phi_c$ is some cutoff at large negative values. We have defined the
finite energy $\epsilon$ by subtracting a divergent piece. This divergence arises
because the string is  stretching all the way to $\phi \to - \infty$.

In the $ \phi \to - \infty$ region,
 \classwkb\ will be a correct approximation for the full wavefunction.
We will check this statement later both on the matrix model side and in the exact
worldsheet analysis. So we can define the scattering amplitude through\foot{
This is a somewhat non-standard definition of the phase. The relation to the more
standard definition can be found in appendix A.
This is a detail  most readers will want to ignore
in a first reading.}
\eqn\phasedef{
\psi = e^{ - i { 1 \over 2 \pi} (\phi - \phi_m)^2} - e^{i \delta}
e^{ i { 1 \over 2 \pi } (\phi - \phi_m)^2}
}
In this definition, $\phi_m$ is just a constant related to the energy $\epsilon$
of the string
through \energ .
In the linear dilaton region we see that the phase \phasedef\ is independent of
the energy since it depends only on the behavior of the wavefunction for
$\phi \sim \phi_m$
and $\phi_m$ drops out of this problem.
In other words, the essential physics of this problem is translation invariant and it
does not matter precisely where the string bounces back.
 We will define the constant relative phase
between the two terms in \phasedef\ in such a way that this constant is zero
\eqn\classphase{
e^{i\delta } =1
}

\subsec{ Strings in (Liouville)$\times$(Time) }

Let us now consider the Liouville case. The general solution of
the  classical Liouville equation can be written as \eqn\gensol{ \mu e^{2 b
\phi} = { 1 \over [ \sum_i f^+_i(b \sigma^+) f_i^-(b \sigma^-) ]^2} }
where $f_i$ obey the conditions \eqn\confi{
\partial^2 f_i - f_i =0~,~~~~~~\epsilon_{ij} f_i \partial f_j =1
} The first comes from \vircond , with $Q = 1/b+b \sim 1/b$, where we assume that
$b$ is small which is the classical limit.  The second equation in \confi\ comes
from the   Liouville  equation. We then conclude that the most general
solution is
\eqn\mostgen{
\mu e^{2 b\phi} =
  { 4 \over
 (  A e^{ b \sigma^+ + b \sigma^-} + B e^{b \sigma^+ - b \sigma^-} + C e^{ -b \sigma^+ +
 b \sigma^-} + D e^{ - b \sigma^+ - b \sigma^- } )^2 }
} where $AD-BC =1$. Some particular interesting cases are $A=D
=1$ $B=C=0$. In this case we get \eqn\solcicl{ \mu e^{2 b \phi} = {
1 \over \cosh^2 b \tau } }
 where $\tau = \sigma^+ + \sigma^- $.
 In this case we can compactify the $\sigma$ direction. This
is the standard solution representing a small string coming in
from infinity.
The spacetime energy of the string is related
to the size of the $\sigma $ circle. If we rescale this circle so that it
has standard length $2\pi$, then we also have to shift $\phi$, since
scale transformations also act on the Liouville field. Then we see
that the maximum value of $\phi$ is related to the energy of the incoming
particle \refsomeone .
 Another interesting solution is $a =
d= \cosh \gamma$ and $b=c = \sinh \gamma$ then the solution is
\eqn\solcomigo{ \mu e^{2 b \phi} = { 1 \over ( \cosh \gamma \cosh
b \tau + \sinh \gamma \cosh b \sigma )^2} }
 This represents the string
coming from infinity and going back to infinity. For
$\alpha =0$ we get \solcicl\ (with $ Q\sim 1/b$).
 In the limit $\alpha \to \infty$ we get back
\solequ . In this case the string ``bounces back'' before it gets to the region
where the Liouville potential is important.

\subsec{Exact worldsheet results for the scattering of long strings}

An interesting observable for these strings is the scattering amplitude for
the strings to go in and come back to infinity. At the classical level (in the $g_s$
expansion)
this is just a phase. We can compute this phase exactly in $\alpha'$ by using
the results for the disk two point functions in \fzzbdy . So we introduce an
FZZT brane \refs{\fzzbdy,\teschnerbdy} that is extended in $\phi$ but with $\mu_B \gg
\sqrt{\mu}$ so that
it dissolves at a very large negative value of $\phi \sim - \log \mu_B$.
Then we consider an open string living on this D-brane. We send in the open string
towards the strong coupling region with high energy $E_0$. When the open string
reaches the region $\phi \sim - \log \mu_B$,   its ends get stuck in this region
while the bulk of the open string stretches as in the solutions we considered above.
Eventually the energy of the stretched string becomes equal to the initial energy.
At this point the string will stretch between $\phi_c \equiv - \log \mu_B$ to
$\phi_m \sim  E_0 \pi - \log \mu_B $ (see the related formula \energ ).
If we take a limit with
\eqn\limit{
 E_0 \to \infty , ~~~~\mu_B \to \infty ~,~~~~~~~~~~ \epsilon = E_0 - { 1 \over \pi} \log
 \mu_B = {\rm finite}
 }
then we get the problem that we are interested in. Namely, we have an infinitely
stretched string that comes in from infinity and goes back to infinity.
We are introducing the FZZT branes and the open strings on them in order to
be able to use the formulas in \fzzbdy\ to compute the amplitude we are interested in.
Starting with the open string two point function in \fzzbdy\ and performing the scaling
limit in \limit , we obtain the scattering phase (see appendix A for details)
\eqn\amplitbos{\eqalign{
\delta(\epsilon) =&   - \int_{-\infty}^{\hat \epsilon } d \epsilon' \,
\left( { \pi \epsilon' \over \tanh \pi \epsilon'} + \pi \epsilon' \right)
\cr
 \hat \epsilon & \equiv
\epsilon + { 1 \over 2 \pi}  \log \mu
}}
The answer is a function of $\hat \epsilon$. This can be understood as follows.
The Liouville potential becomes important at $\phi_L \sim - \half \log \mu$.
The energy $\epsilon$ is related to the point were the string
would stretch to a maximum in the absence of Liouville potential  by \energ .
So the amplitude is just a function of $\phi_m - \phi_L$, which is proportional to
$\hat \epsilon$.
We see that the phase  $\delta(\epsilon) \to 0$ for $\epsilon \to -\infty$, so
that we recover the linear dilaton result \classphase .
In fact, the asymptotic behavior of this phase is
\eqn\limbeh{
\delta(\epsilon)    \sim   (\hat \epsilon - { 1 \over 2 \pi} )
e^{ 2 \pi \hat \epsilon} =
 (\hat \epsilon - { 1 \over 2 \pi} ) \mu e^{ 2 \pi \epsilon} ~,~~~~~~~~~~~{\rm
 for} ~~~~~ \epsilon \ll \log \mu
}
We see that the leading corrections to the $\epsilon \to -\infty$ answer has the
form we expect. To leading order we get that the phase is zero, in agreement with
\classphase , and the first correction can be interpreted as resulting from the insertion
of a cosmological constant operator.  Indeed, in this case the string bounces back
before it gets to the region where the Liouville potential is important.
On the other hand, the high energy behavior is
\eqn\limhe{
 \delta(\epsilon)   \sim  - \pi \hat \epsilon^2 +
 ( \hat \epsilon + { 1 \over 2 \pi})  e^{ - 2 \pi \hat \epsilon}
    ~,~~~~~~~~~{\rm for}~~~ \epsilon \gg \log \mu
}
The leading result in \limhe\  can also be easily understood.
We can think of the Liouville potential as a sharp wall at $\phi_L = - \half \log \mu$.
This imposes a boundary condition on the wavefunction at this point, say
  $\psi(\phi_L) =0$.
  From \phasedef\ we see that this implies that
\eqn\largee{
\delta \sim - { 1 \over \pi} ( \phi_L - \phi_m)^2 = - \pi \hat \epsilon^2
}
which is just the leading behaviour of \limhe .
Note that if the energy gets extremely large, $\epsilon \sim \mu$, then we expect
that this approximation would fail, since that is enough energy to create a ZZ
brane. But for large $\mu$ there is a large range of energies, $\mu \gg \epsilon \gg
\log \mu$,  where we can
trust \limhe .

\subsec{ Worldsheet formulas for the two dimensional type 0 superstring}

For the type 0  superstring formulas we set $\alpha'=\half$. By starting with the
exact Liouville formulas in \fukuda\  and taking a limit where
$\mu_B \to \infty$  we find two possible results (see appendix A)
\eqn\twopos{
\delta_+( \epsilon) = - 2 \int_{-\infty}^{ \hat \epsilon \over 2}  d\epsilon'
 \pi \epsilon' ({  \tanh \epsilon' \pi  } +1) ~,~~~~~~~~~~~~~
  \delta_-( \epsilon) = - 2 \int_{-\infty}^{ \hat \epsilon \over 2}  d\epsilon'
 \pi \epsilon' ({ 1 \over \tanh \epsilon' \pi  } +1)
}
with
\eqn\defeph{
\hat \epsilon \equiv \epsilon + { 1 \over \pi } \log \mu
}
The first thing we need to understand is why we get two different results.
Let us first look at the behavior of \twopos\ for $\hat \epsilon \ll 0$.
To leading order both results give us that $\delta = 0$, which agrees with the
result one expects in the linear dilaton region. The first correction is
\eqn\leadob{
\delta_\pm \sim   \mp ( \hat \epsilon - { 1 \over \pi}) e^{\pi  \hat\epsilon} = \mp
( \hat \epsilon -{ 1 \over \pi} ) \mu e^{ \pi  \epsilon}  ~,~~~~~~~~\hat \epsilon \ll 0
}
So the only difference in the two results is the overall sign.
In the linear dilaton region the field $\phi$ is given by \solequ . In a
gauge where the worldsheet fermion $\psi^0=0$ we also need to set to zero the
supercharge in the $\phi$ direction. This will lead to the equation
\eqn\superch{
 \psi^\phi_+ \partial_+ \phi - Q \partial_+ \psi^\phi_+ =0
 }
 and a similar equation for left movers. The only solution to \superch\ is
 \eqn\solzm{
 \psi^\phi_+ = { 1 \over \cosh {\sigma^+ \over Q} } \epsilon_0
 }
 where $\epsilon_0$ is a constant. It can be checked that this mode is in the
 cohomology, namely it is not given by acting with the supercharge on the original
 configuration \solequ .
Upon quantization, the left and right moving modes $\epsilon_0^{l,r}$ together lead
to a two dimensional Hilbert space ($\epsilon_0^{l,r} \sim \sigma^{1,2}$, where
$\sigma^i$ are Pauli matrices).
The cosmological constant term $\mu \psi^\phi_+ \psi^\phi_-e^{ 2\phi}$   reduces
to $ \sigma^3 \mu e^{2 \phi}$ when it is projected onto the two dimensional space
Hilbert space. So we see that we will get two different answers. When the scattering
occurs in the large negative $\phi$ region the leading correction comes from inserting
just one cosmological constant operator and this leads to the two possible signs in
\leadob .

The results in    \twopos\ were actually obtained
by scattering NS open strings on an FZZT brane with
$\eta = \pm 1$. Recall that $\eta$ is a sign which relates the left moving to the
right moving supercharge at the boundary of the worldsheet.
We are using the conventions of \fukuda\ where the ZZ brane has
$\eta = -1$.
The sign of $\eta$ is telling us the relative sign between the left and right moving
fermion. Since the cosmological constant term goes as $\mu \psi_L \psi_R e^{ 2 \phi}$,
the relative sign between the left and right moving fermion determines the sign
of the first correction. The precise boundary conditions at the end of the string
should impose that only one of the two states associated to the fermion zero modes we
discussed above are kept. We have not worked out the details of how this happens, but
it follows from the results in \fukuda .

The results are basically the same in type 0A and 0B theories, though some of
the detailed properties of the FZZT branes are different.

We can also consider the scattering of Ramond open strings ending on  FZZT branes with
different $\eta$. In the $\mu_B \to \infty$ limit that we are considering we find
results as in \twopos.

\newsec{ Non-singlets in the matrix model}

In this section we consider the $SU(N)$ gauged quantum mechanics of a single
hermitian matrix $\Phi$ with the insertion of a Wilson line in the representation
${\cal R}$
\eqn\defmod{
 Z = \int {\cal D} \Phi {\cal D}A e^{ \int dt Tr [ (D_0 \Phi)^2 - V(\Phi)] }
 Tr_{\cal R} e^{ i \int A }
 }
When we analyze this model we can equivalently set $A_0=0$ and consider only states
in the Hilbert space of the matrix $\Phi$ that transform in the representation
${\cal R}^\dagger$. From now on we will replace ${\cal R} \to {\cal R}^\dagger $ for
notational convenience.
This problem was considered in
\refs{\mo,\gkvortex,\boulkaz,\minahanpol,\polreview,\dasgupta}
  \foot{One minor difference
with some of these papers is that here we do not include a factor proportional
to the dimension of the representation in the partition function. }.
Of course, the representation ${\cal R}$ should be such that it
can be constructed by taking multiple products of the adjoint representation. So
 the representation should be invariant under the center of $SU(N)$.
We are going to be interested in taking the large $N$ limit and we want to keep
the representation ${\cal R}$ ``fixed". In this limit it is useful to think
of the Young diagram for
representation as made up with some boxes and ``anti-boxes'', see figure three.
We have the constraint that the number of boxes and anti-boxes should be
equal. But the Young diagram for the boxes need not be equal to the Young diagram
for anti-boxes. In other words the representation need not be self
conjugate \boulkaz \foot{Only self conjugate representations were considered in  \gkvortex .
 }.

\ifig\young{ Young diagram of a possible representation. In this case we have three
boxes and three anti-boxes.   }
{\epsfxsize 1.5 in\epsfbox{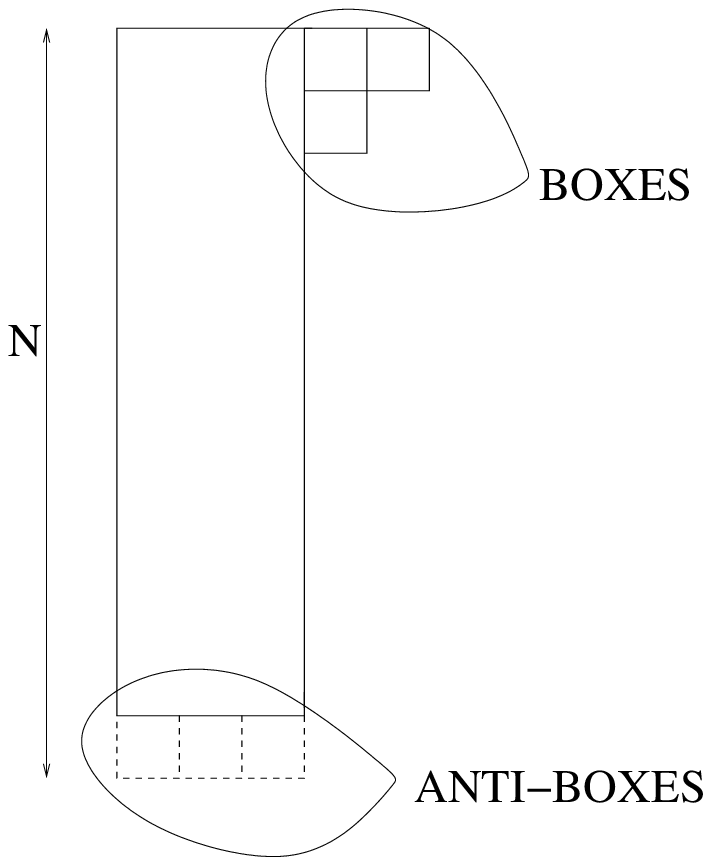}}

It is convenient to diagonalize the matrix $\Phi = diag( \lambda_1 , \cdots, \lambda_M)$.
This gauge choice breaks $SU(N)$ to $U(1)^{N-1}$.
Then we can write the wavefunction as $\psi_\alpha(\lambda_i )$ where $\alpha$ is
an index in the representation ${\cal R}$, which labels the states with zero
weight in this representation, namely states with zero $U(1)^{N-1}$ charges.
The restriction to zero weight states comes
from integrating out the diagonal part of the gauge field $A_0$.
The hamiltonian is
\eqn\hamiltred{
H =  \left[ \sum_{i=1}^n  - \half { \partial^2 \over \partial \lambda_i^2} -
\half \lambda_i^2  + { 1 \over 2} \sum_{i \not =  j }
{ T^{\cal R}_{ij} T^{\cal R}_{ji} \over ( \lambda_i - \lambda_j)^2 } \right] P_0
}
where $P_0$ is the projector on to states in the representation ${\cal R}$ with
zero $U(1)^{N-1}$ charges.
From now on we will drop it. $T^{\cal R}_{ij}$ are the matrices
corresponding to the $ij$ generator of $SU(N)$ in the representation ${\cal R}$.

\subsec{ The adjoint representation}

This case was considered for a general potential in \mo .
Here we review their procedure and
we will further explore it in the case of the inverted harmonic oscillator, where
some of its features were discussed in \gkvortex .
The adjoint representation is just given by a matrix $W_{i}^j$. The states with
zero weight are  diagonal matrices $W_{i}^j = w_i \delta_{i}^j$, with
$\sum_i w_i =0$. We see that the projection to zero weight states reduced the
number of states from $N^2-1$ to $N-1$.

We will now treat the interaction in \hamiltred\ perturbatively. To zeroth order
we neglect it and the Hamiltonian is just that of free fermions, with the usual
ground state. This ground state is degenerate since we can choose any state $W$ in
the representation. To first order the interaction term in \hamiltred\
removes this degeneracy. We can think of the representation part of the
wavefunction as given by a state vector $|W\rangle = \sum_i  w_i |i \rangle \times | i \rangle $
Then the interaction part of the Hamiltonian acts as
\eqn\intactp{
 H_{int} |W \rangle = \sum_{i=1}^N \left(
 \sum_{j=1, ~j\not= i}^N { (w_{i} - w_j) \over (\lambda_i - \lambda_j)^2 } \right)
 |i\rangle
 \times |i \rangle
 }
The interaction Hamiltonian is positive definite. This is most easily
 seen by computing $\langle W |H_{int}| W \rangle$.
After introducing the eigenvalue density in the standard way and writing
$w_i \to w(\lambda)$
we end up with the following eigenvalue problem
\eqn\eigenv{
E w(\lambda) = \int d\lambda' \rho(\lambda') {w(\lambda) - w(\lambda') \over
(\lambda - \lambda')^2 }
}
It looks like the lowest energy state is $w(\lambda) =$constant. However, this
state  is not in the adjoint representation which is constrained
to have $\int d\lambda \rho(\lambda) w(\lambda) =0$. Some approximate solutions
to this problem can be found in \mo .

Here we are interested in understanding the inverted Harmonic oscillator potential.
In this case it is more convenient to introduce a slightly different variable
\eqn\newvar{
h(\lambda) = w(\lambda) \rho(\lambda)
}
In terms of this variable the eigenvalue problem we need to solve becomes
\eqn\weha{\eqalign{
 E h(\lambda) = & - \rho(\lambda) \int { d \lambda'  } {
 1  \over (\lambda - \lambda')^2 }
   h(\lambda')
 + v(\lambda) h(\lambda)
\cr
v(\lambda)   & \equiv   \int d\lambda'  {   \rho(\lambda') \over
(\lambda - \lambda')^2 } \cr
 1 = & \int { d \lambda \over \rho(\lambda)}  h^2(\lambda)
 }}
 where we have also included the form of the normalization condition\foot{ Note
 that  $v(\lambda)$ in \weha\ is not the same as $V(\lambda)$ in the original
 matrix model Lagrangian.}.
In all integrals we should take the principal value, which means that we subtract all
divergent terms when $\lambda' \to \lambda$.
The first term \weha\ can be viewed as a kinetic term and
the second as a potential. This potential contains a constant divergent term
plus a finite term. Let us first understand the problem in the asymptotic region
where $\lambda \gg \sqrt{\mu}$, we will later discuss the more general case.
In the asymptotic
region we can approximate $\rho(\lambda) = { 1 \over \pi}  \lambda$ and define
$ \tau \equiv \log \lambda $. We  obtain
\eqn\masste{
v(\lambda) =   { 1 \over \pi} \int_0^{\lambda_{c}}
 d\lambda' { \lambda' \over (\lambda' - \lambda)^2 } \sim
{ 1 \over \pi} ( \tau_{c} - \tau -1 )
}
where $\tau_c$ is some large cutoff value of $\tau$. The divergence comes
from the eigenvalues that are far away, at large $\lambda'$. The region
$\lambda' \sim \lambda$ does not lead to a divergence because we are using a
principal value prescription.
So, up to the divergent term we find a linear potential that is pushing the
particle to the large $\tau $ region.
Similarly we can study the kinetic term in the asymptotic region and we find
that
  the normalization condition and the  kinetic term, $K$,  become
\eqn\simplbec{
\int  d\tau   h^2(\tau) =1~,~~~~~~~~~
(K h )(\tau)= -  { 1 \over \pi}    \int {  d \tau'}{  h(\tau')
\over  (2 \sinh({\tau - \tau' \over 2}))^2 }
}
Since the kinetic term became translational invariant it is reasonable
 to solve the problem using plane waves. Computing the kinetic term in
Fourier space we find
\eqn\plw{ K(k) = -{ 1 \over \pi} \int_{-\infty}^{\infty} d\tau { \cos k \tau
\over 4 \sinh^2 {\tau \over 2}   }=  {  k  \over \tanh( \pi k) }
}
where we take the principal part of the integral (i.e. we disregard terms that diverge
as $\tau \to 0$). At large $k$ we have a
relativistic relation
\eqn\relat{
E \sim |k| ~~~~{\rm for}~~|k|\gg 1
}

So we see that the Hilbert space spanned by the vector $w_i$ can be thought of as
the Hilbert space of the particle that lives at the tip of the string.
The string provides
the linear potential \masste . It is interesting to see how the dynamics of this
extra particle is generated
 out of the interaction with the background fermions.
 The coefficients in \masste\ and \plw\ match
precisely with our expectations based on the string theory discussion after
we identify $ \tau \sim - \phi$, where $\phi$ is the Liouville direction.
This is a standard identification valid for large
values of $\tau$ \refsomeone . In fact, the divergence in \masste\ is related
to the fact that the string is stretched all the way to $\phi = - \infty$.
We can solve the complete quantum scattering problem in this asymptotic
region and one finds a result that agrees with \classphase , see appendix B.

A few remarks are in order. First, note that, after subtracting the infinite
additive constant \masste , the spectrum is {\it unbounded below }\foot{
The authors of \gkvortex\ stated that the spectrum has a gap
of order $\log \Lambda/\mu$ (which is the infinite subtraction we are performing)
and then a
continuum spectrum similar to that of the
free fermions. What is important is that this continuous
spectrum is unbounded below.}.
 This has some important consequences when
we examine the physics of this model.
Second, the infinite constant that we are subtracting behaves as
\eqn\constsub{
E_{div} \sim  { 1 \over \pi} \tau_{c} = - { 1 \over \pi } \phi_{c}
}
There is precise agreement in the coefficient of this divergent term between
the matrix model and  string theory.

Now let us come back to the general problem in \weha\ and include the $\mu$ dependence.
It is convenient to define the $\tau $  variable through
$\lambda = \sqrt{ 2 \mu} \cosh \tau$. Then $\rho(\lambda) = \sqrt{ 2 \mu} \sinh \tau$.
We define $h(-\tau) \equiv -  h(\tau)$ for $\tau >0$. The sign in this definition is
motivated by the fact that we expect a boundary condition $h(\tau =0) =0$, since
at $\tau=0$ we run out of fermions which could carry the adjoint
indices. With this definition of $h$ for negative values of $\tau$ we can extend
the range of the integral over the whole $\tau$ axis and obtain a simple expression
for the kinetic term. Namely, the same as the one we had in \simplbec .
We also subtract the divergent piece of the energy. The eigenvalue problem then becomes
\eqn\newform{\eqalign{
 \hat \epsilon h(\tau) = & - {1 \over \pi} \int_{-\infty}^\infty d\tau' {
  h(\tau') \over 4 \sinh^2{\tau - \tau' \over 2} } + \hat v(\tau) h(\tau)
\cr
\hat v(\tau)  =&  - { 1 \over \pi}  { \tau \over \tanh \tau }
}}
with
\eqn\defofe{
\hat \epsilon =  E - {1 \over \pi} \tau_c + {1 \over \pi}  =
 E + { 1 \over \pi } \log\lambda_c + { 1 \over 2\pi}
\log \mu + {\rm const}  = \epsilon + { 1 \over 2 \pi } \log \mu
}
where $\epsilon$ should be thought of as the renormalized energy since it is
the one defined with a $\mu$ independent subtraction procedure. We have absorbed some
constants in this subtraction in order to make formulas look simpler.
Note that $\mu$ has completely disappeared from \newform , which agrees with
the $\mu$ dependence of the string theory result \amplitbos . The energy $\epsilon$
defined in \defofe\ agrees with the renormalized energy defined in \energ\ or \limit ,
up to an additive  numerical constant which we have fixed by comparing the matrix model and
the string theory answers. In summary, the energy $\epsilon$ which appears  in
\defofe\ should be directly identified with $\epsilon$ in \amplitbos .

We have been unable to solve the problem of computing the phase directly from
\newform . Nevertheless we conjecture that the scattering phase that comes
out of it is given by
\eqn\matchres{
 \delta(\hat \epsilon) = - \int^{\hat \epsilon}_{-\infty} \left(
  { \pi \epsilon' \over \tanh \pi \epsilon'} + \pi \epsilon' \right)
 }
 with a definition of the scattering phase analogous to \phasedef .
We have computed explicitly the two asymptotic limits, $\epsilon \to \pm \infty$,
 as well as
the first exponential subleading correction in each of these asymptotic regions and
checked that they match \limbeh \limhe . See
appendix B for details. It
seems very likely
 that there is a simple  way to solve this problem exactly to derive \matchres .

\subsec{ The vortex anti-vortex correlator}

So far we have been discussing the Lorentzian theory. Let us now consider the
Euclidean theory with euclidean time compactified on a circle of radius $R$.
Thus we compute the thermal partition function for the matrix model in the adjoint
representation.
 On the worldsheet it is easy to compute it.
 It is given by the two point function for two winding strings. Let us first
 perform an approximate qualitative analysis.
The Liouville correlation function
involves an integral of the from
\eqn\integr{
 \int d \phi e^{ - \mu e^{ 2 \phi} } e^{ 2 ( 2 - R) \phi} e^{ -  4 \phi}
 }
 The last term comes from the dilaton. We see that this diverges at weak coupling
 as $\int d \phi e^{ - 2 R \phi}$. This divergence, which
 has a positive coefficient,
 is related to the energy of the stretched string. In other words,
  expect a divergence
 of the form $\int d \epsilon e^{ - (2 \pi R) \epsilon}$ for $\epsilon \to -\infty$.
  This divergence
 is related to the fact that, after the subtraction, the spectrum of the matrix model
 Hamiltonian is unbounded below. This divergence is independent of $\mu$.
 If we concentrate only on $\mu$ dependent terms, then we can get a finite answer.
We can compute this
  by taking a derivative
 with respect to $\mu$ of \integr . This then gives a convergent
 integral for $ R<1$
 \eqn\integweu{
\partial_\mu \langle 2pt \rangle \sim
 - \int d \phi e^{ - \mu e^{ 2 \phi} } e^{ 2 \phi (1-R) } \sim  - \mu^{R -1} \Gamma( 1-R)
}
This has just been an approximation to the true answer. The correct answer is computed
using the exact expression for the Liouville three point function in \zz . After taking
 the limit $b \to 1$ in the formulas in \zz\ we obtain
\eqn\cftres{
\partial_{\mu} \langle 2pt \rangle_{Liouville} = - \langle 3 pt \rangle_{Liouville} = -
{ { \mu}^{ 1-R}  \over \pi} R^2 { \Gamma(-R)^2 \over
\Gamma( R)^2 }
}

This implies that the two point function in the full string theory
 is then given by
\eqn\twopotex{
\langle 2pt \rangle = - 2   R^2 {  \mu }^{ R} { \Gamma(-R)^2 \over
\Gamma(R)^2 } \sim { Z_{adj} \over Z_{singlet} }
}
up to a (positive) numerical constant. Going from \cftres\ to \twopotex\
 we loose one factor of $R$ from integrating \cftres\ and we gain one
factor of $(2 \pi R)$ from the $X$ path integral. Notice that this formula
is consistent with the statement that the free energy in the adjoint
differs from the singlet free energy  by a logarithmic term
\eqn\freen{
F_{adj} \sim F_{sing} - { 1 \over 2 \pi  } \log \mu
}
Note that the coefficient of the logarithmic term agrees precisely with the matrix
model\foot{ The computations in \refs{\gkvortex,\boulkaz} were   off by
a factor of two. },
due to the agreement in the divergent parts of the energy that we checked
above, see \constsub .
 The sign in \twopotex\ seems puzzling.
Naively one would have expected the partition function in the adjoint representation
to be
positive. What happens is that we have made an infinite subtraction due to
the fact that the spectrum is unbounded below.

Now we would like to compare \twopotex\ with the matrix model answer. We
are interested in computing the matrix model partition function in the adjoint
representation.
The matrix model answer depends on the scattering phase
\matchres .
We are then lead to a partition function for the adjoint of the form
$Z_{adj} \sim  Z_{sing} Z_W$, where $Z_{sing}$ is the singlet partition function and
$Z_W$ is the partition function for the effective particle we are considering now.
\eqn\resinteg{\eqalign{
Z_W =&  \int d\epsilon { \partial_\epsilon \delta(\epsilon ) \over 2 \pi}
e^{ - 2 \pi R \epsilon}
= - \mu^R \int_{-\infty}^\infty   d\hat \epsilon  \half \hat
\epsilon ({ 1 \over \tanh \pi \hat \epsilon} +1)
   e^{ - 2 \pi R \hat \epsilon}  = -{ 1 \over 4 \sin^2 \pi R } \mu^R = \cr
   &
    =
    - { \mu^R \over 4 \pi^2} \Gamma(-R)^2 R^2 \Gamma(R)^2
}}
We see that we reproduce \twopotex\ up to factors that are analytic in $R$ for positive $R$.
The difference between \cftres\ and \resinteg\ is reminiscent of
the leg factors that appear
when one compares scattering computations in string theory and the matrix model
\refsomeone\ but it is different in detail\foot{The leg factor that we need in this case
to match the matrix model answer \resinteg\ to the string theory answer \twopotex\
 is $ \Gamma(R)^{-2}$ for each
vertex operator, which looks
 different than the momentum leg factors, which are
  $\Gamma(-|p|) \over \Gamma(|p|)$ (and we would set $p=R$).}.
\resinteg\ is the correct normalization for the partition function of the adjoint since
we constructed it explicitly as a trace over the Hilbert space. This normalization
  still has an ambiguity due to the precise constant in the subtraction of the energy. This
means that
that we can replace $\mu \to \mu c$ where $c$ is an $R$ independent numerical constant.
Notice that the poles at integer values of $R$ in \resinteg\ arise because
when $R$ is an integer we have a new divergence in the computation of the free
energy since the integrand in \resinteg\ does not decay fast enough at infinity. This
implies that a new subtraction becomes necessary. The coefficients of these divergent
terms are integer powers of $\mu$.

We now consider the problem from a slightly different point of view.
Let us consider an FZZT brane extended in time and compute its cylinder diagram.
This diagram contains a contribution from the exchange of closed strings
that are wound in the time direction. We can pick out the contribution from
strings that wind precisely once as follows.
The cylinder diagram is (up to a positive normalization constant)
\eqn\cylbos{
Z_{cyl} \sim  2 \pi R \sum_n \int_0^\infty dE { \cos^2(\sigma \pi E) \over \sinh^2\pi E } { 1
\over E^2 + n^2 R^2 }
}
The integer $n$ is summing over the closed string exchanges with strings that are wound
$n$ times, so we will be interested in the term with $n=1$.
We disregard the pole at $E=0$ and all other terms that are $\mu$ independent.
 We can rewrite the integral in \cylbos\ as an
integral from $-\infty $ to $\infty$. Then we split the cosine into $e^{ 2 i \sigma E}$
and its complex conjugate. Both terms give the same answer.
 For large $\sigma$ it  is convenient to shift the contour
of the first to $E \to i \infty$. In this process we pick up poles at
$E=i m$ and $E= i n R$. If $ 0 < R<1 $ the leading pole appears at
$E = i R$ and its residue gives
\eqn\respol{
Z = -2 \pi^2  e^{ - 2 \sigma \pi R }{ 1 \over \sin^2 \pi R} \sim
 -  {  \mu}^R (\mu_B)^{-2 R} { 1 \over \sin^2 \pi R}
 }
Note that also from this point of view we get a negative sign, in agreement
with the worldsheet computation. Note that here, also, we started with \cylbos\ which
is manifestly positive and, due to a subtraction, we ended up with a negative answer.
The precise agreement between \respol\ and \resinteg\ is simply a consistency check,
since it is related to the fact that we assumed that the scattering phase in the
matrix model agrees with the scattering phase in string theory.

 We could take a scaling limit where we put $N_f$ FZZT branes and then take
\eqn\limscl{
N_F , \mu_B \to \infty, ~~~~~~~~~~ \hat \lambda \equiv  N_f \mu_B^{-R} = {\rm fixed}
}
 \respol\  gives a finite source for the winding
modes. This procedure would give a source only for the first winding mode.
We should then identify $\hat \lambda$ with the term in the  euclidean action
considered in \kkk
\eqn\kkpart{
S = S_0 + \hat \lambda Tr\left(e^{\int A}\right) + \hat \lambda Tr\left(e^{\int A}
\right)^\dagger
}
Thinking of \kkpart\ as a limit of \limscl\
has the advantage that we start from system with
a clear Lorentzian interpretation, at least on the string theory side, which
in the limit \limscl\ leads us to consider the  partition function for the action
 \kkpart\ studied
in \kkk . So the two dimensional
string theory in the presence of $N_f$ FZZT branes might be a good starting
point for thinking about the Lorentzian version of the two dimensional
black hole \twodbh .

\subsec{All orders result via T-duality}

We expect the all orders partition function for the adjoint representation
 will be given the correlator
of two strings with winding numbers one and minus one. We can compute this in the
following indirect way. From the string theory point of view
 we can assume that T-duality is a good symmetry.
  On the T-dual side, this translates into a computation
for momentum correlators. We can compute these using the matrix model fermions
of the T-dual model, which are free since we are simply computing correlators of
momentum operators. This is just a computation in the singlet sector and
can be done using the general formulas in \greg . The matrix model fermions of the
T-dual model should not be confused with the original matrix model fermions. T-duality
is not an obvious symmetry of the matrix model.
We find, see details in appendix D,
\eqn\correl{
Z \sim  Z_{sing} \left( {
\Gamma( \half + i \mu R + {R \over 2} ) \Gamma( \half -i \mu R + {R \over 2})
\over
\Gamma( \half - i \mu R - {R \over 2} ) \Gamma( \half +i \mu R - {R \over 2}) }
\right)^{1/2}
}
up to non perturbative terms in $1 \over \mu$, which is us much precision
as we can hope for the  bosonic string case. The correlator \correl\ is simply the momentum correlator
in the T dual matrix model written in terms of the original variables.
The result \correl\ differs from the result in \boulkaz\ by the overall power (instead
of $\half$ in the exponent they had ${ 1 \over 4}$ \foot{ Equivalently,
 \boulkaz\ obtain only the numerator in \correl\ , which is equivalent to changing
the overall power up to terms that are analytic in $\mu$
and non-perturbative in $1/\mu$.}).
On the T-dual matrix model we know that there are leg factors that need to be taken
into account. Notice also  that \twopotex\ differs from \resinteg\ by another leg
factor. We do not understand the origin of these leg factors in a clear way.
So assuming that they will work out as in the usual singlet case we conclude that
the all orders partition function in the singlet sector should be given by
the the $1/\mu$ expansion of
\eqn\correlf{\eqalign{
Z \sim  &  - { 1 \over 4 \sin^2 \pi R}Z_{sing} \left( {
\Gamma( \half + i \mu R + {R \over 2} ) \Gamma( \half -i \mu R + {R \over 2})
\over
\Gamma( \half - i \mu R - {R \over 2} ) \Gamma( \half +i \mu R - {R \over 2}) }
\right)^{1/2}
\cr
Z \sim  & - Z_{sing} \times { \mu^R \over 4 \sin^2 \pi R} \left( 1 + ({R \over 24} - { 1 \over 24 R})
{ 1 \over \mu^2}  + \cdots \right)
}}
where we introduced an $R$ dependent factor in order to match
the leading order answer
to \resinteg . These could come from leg factors which we do not fully understand.

\subsec{More general representations}

We saw that if we consider the adjoint representation we basically get one
more particle. If we have a representation with $n$ boxes and $n$ antiboxes, then
to leading order we will get a dilute gas of $n$ particles which will have
the same properties as the particle we had for the adjoint representation.
The string theory dual will thus consist of $n$ folded strings that come from the
weak coupling region. The tip of these strings comes in, into the strong coupling region,
and then goes back out to the weak coupling region.
The details about the representation are related to symmetry properties of these
particles, see \minahanpol\ for further discussion.

An interesting class of representations
are those which are totally anti-symmetric in boxes and totally
anti-symmetric in antiboxes.
See figure 4 (c).
In this case we can map the problem into an $SU(2)$ spin chain  \minahanpol .

\ifig\onerow{ Young diagrams that give rise to $SL(2)$ spin chains (a), SL(2)
spin chains (c) and the supersymmetric Calogero model (b).}
{\epsfxsize 1.0 in\epsfbox{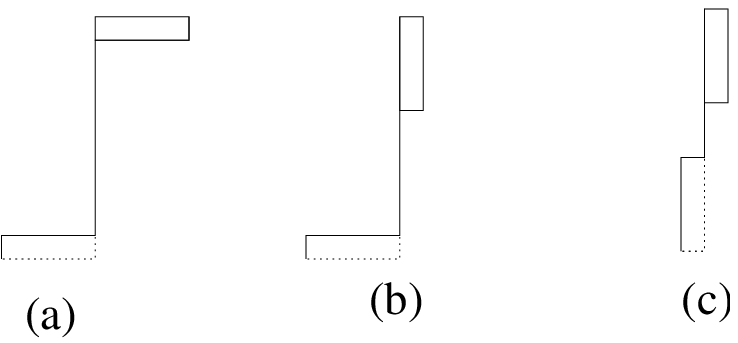}}

We can endow each matrix eigenvalue, which is a fermion, with an $SU(2)$ spin and
an interaction which is non-zero and goes as $1/(\lambda - \lambda')^2 $
if the spins are in a singlet while it is zero if
they are in the triplet of $SU(2)$. Then we see the singlet sector corresponds to
the case that all spins are up (we have maximal spin under global $SU(2)$). The adjoint
corresponds to flipping one spin, and similarly a representation with $n$ symmetric
boxes and $n$ symmetric anti-boxes corresponds to flipping $n$ spins (most properly
we should say that the total spin is $J^3 = N/2 - n/2$).

Another interesting set of representations are those completely symmetric in
boxes and anti-boxes, see figure 4 (a). In this case we get an $SL(2)$ spin chain.

Finally we could consider representations which are anti-symmetric in boxes and  symmetric
in anti-boxes, see figure 4 (b).
In this case we also get some particles that move on the fermi sea, but
they are fermionic. In fact, we can think of each fermion of the matrix model as
endowed with an extra fermionic degree of freedom. The final theory is the supersymmetric
Calogero-Moser model \ref\DabholkarTE{
  A.~Dabholkar,
  ``Fermions and nonperturbative supersymmetry breaking in the one-dimensional
  superstring,''
  Nucl.\ Phys.\ B {\bf 368}, 293 (1992).
}\ref\FreedmanGD{
  D.~Z.~Freedman and P.~F.~Mende,
  ``An Exactly Solvable N Particle System In Supersymmetric Quantum
  Mechanics,''
  Nucl.\ Phys.\ B {\bf 344}, 317 (1990).
}\ref\brinkva{ L.~Brink, T.~H.~Hansson, S.~Konstein and M.~A.~Vasiliev,
  ``The Calogero model: Anyonic representation, fermionic extension and
  supersymmetry,''
  Nucl.\ Phys.\ B {\bf 401}, 591 (1993)
  [arXiv:hep-th/9302023].
}\polreview \ref\BergshoeffDD{
  E.~Bergshoeff and M.~A.~Vasiliev,
  ``The Calogero model and the Virasoro symmetry,''
  Int.\ J.\ Mod.\ Phys.\ A {\bf 10}, 3477 (1995)
  [arXiv:hep-th/9411093].
}\ref\desrosiers{
  P.~Desrosiers, L.~Lapointe and P.~Mathieu,
  ``Generalized Hermite polynomials in superspace as eigenfunctions of the
  supersymmetric rational CMS model,''
  Nucl.\ Phys.\ B {\bf 674}, 615 (2003)
  [arXiv:hep-th/0305038].
   P.~Desrosiers, L.~Lapointe and P.~Mathieu,
  ``Supersymmetric Calogero-Moser-Sutherland models: superintegrability
  structure and eigenfunctions,''
  arXiv:hep-th/0210190.
  }\ref\GhoshHY{
  P.~K.~Ghosh,
  ``Super-Calogero model with OSp(2$|$2) supersymmetry: Is the construction
  unique?,''
  Nucl.\ Phys.\ B {\bf 681}, 359 (2004)
  [arXiv:hep-th/0309183].
}\ref\RodriguesBY{
  J.~P.~Rodrigues and A.~J.~van Tonder,
  ``Marinari-Parisi and supersymmetric collective field theory,''
  Int.\ J.\ Mod.\ Phys.\ A {\bf 8}, 2517 (1993)
  [arXiv:hep-th/9204061].
}\ref\JevickiYK{
  A.~Jevicki and J.~P.~Rodrigues,
  ``Supersymmetric collective field theory,''
  Phys.\ Lett.\ B {\bf 268}, 53 (1991).
}\ref\vanTonderVC{
  A.~J.~van Tonder,
  ``A Continuum description of superCalogero models,''
  arXiv:hep-th/9204034.
}, with an inverted harmonic oscillator potential,
with a total $U(1)$ charge which is equal to $N-n$ where $n$
is the number of boxes.
The super-calogero model was proposed in  \ref\verlinde{ H.~Verlinde,
  ``Superstrings on AdS(2) and superconformal matrix quantum mechanics,''
  arXiv:hep-th/0403024.
} as a dual of string theory on $AdS_2$.

\subsec{ Two sided (or 0B)  matrix model}

In this subsection we consider a hermitian matrix model but we fill the two sides
of the inverted harmonic oscillator potential. This is also called the 0B matrix
model, since it is dual to type 0B string theory \refs{\tato,\chatone}.
Since we are choosing $\alpha'={1 \over 2}$ the matrix model potential is exactly
the same as in the bosonic string theory case. So,
up to equation \weha , we have the same discussion as in the bosonic string case.
The new feature is that we need to consider both positive and negative
values of $\lambda$. So we write $\lambda = \pm \sqrt{ 2 \mu} \cosh \tau$
on these two sides. We also define the functions $h_{l,r}(\tau)$ which are defined
on the left and right side of the inverted harmonic oscillator potential.
Here we will
assume that the Fermi level is below the barrier.
As in the bosonic case we extend the range of the  functions
$h_{l,r}( -\tau) = - h_{l,r}(\tau)$ for $\tau \geq 0$. Then
the eigenvalue equations become
\eqn\findhpl{
E h_r(\tau) =  - { 1 \over \pi} \int_{-\infty}^{\infty} d\tau'
\left[{h_r(\tau') \over 4 \sinh^2{\tau - \tau' \over 2} } +
{ h_l(\tau') \over 4 \cosh^2 { \tau - \tau' \over 2} }\right]
+  v(\tau) h_r(\tau)
}
with
\eqn\potenpm{
 v (\tau) \equiv { 2 \over \pi} (\tau_{max} -1 - { \tau \over \tanh \tau}  )
}
We have a similar equation with $r \leftrightarrow l$.
Due to the $Z_2$  symmetry of the problem
we can define $h_+ $ and $h_-$ where $h_+$ is such that
$h_r(\tau)  =  h_l(\tau) $ and $h_-$ is such that $h_r=  - h_l $.
The two kinetic terms that appear in \findhpl\ can be diagonalized to
\eqn\kinen{\eqalign{
\hat \epsilon h = &  K_\pm h + \hat v  h
\cr
K_\pm &=  {  k \over \tanh \pi k}  \mp { k \over \sinh \pi k}  =
k \left(\tanh { \pi k \over 2} \right)^{\pm 1}
\cr
 \hat v & =- { 2 \over \pi} { \tau \over \tanh \tau}
 \cr
\hat \epsilon = & E - { 2 \over \pi } (\tau_c -1) =
E + { 2 \over \pi} \phi_c + { 1 \over
\pi } \log \mu + {\rm const} = \epsilon + {1 \over   \pi } \log \mu
}}
where we have subtracted the divergent terms from the potential.
We now see that the divergent energy is twice what we had in the bosonic case, which
is what we expect after we identify $ \sqrt{\mu} e^\tau \sim e^{-\phi}$
and use $\alpha' =\half $.
  Note that the term involving $\log \mu$ in the definition
of $\hat \epsilon$ has a different coefficient than in the bosonic string
 because of the factor of two in $\alpha'$.
  In this units the Liouville cosmological constant term has the form
 $\mu \psi \bar \psi e^{2 \phi}$.
As in the bosonic case we were not able to solve the eigenvalue problem \kinen\ directly.
We have checked, however, that in the two asymptotic limits $\epsilon \to \pm \infty$
the leading and first subleading contributions computed from \kinen\ match
\leadob .

\subsec{The vortex anti-vortex correlator  for the superstring}

In the type 0B superstring we have only one kind of winding string in the NS sector.
As in the bosonic string we can compute the two point function using the
three point function in \fukuda \foot{It is the one called $C_3(\alpha_i)$ in
\fukuda .}. This gives
\eqn\twons{\eqalign{
- \partial_\mu \langle e^{ 2(1-R) \phi} e^{2(1-R) \phi} \rangle_{Liouv} = &
 ~ i \langle \psi \bar \psi e^{ 2 \phi} e^{ 2(1-R) \phi} e^{2(1-R) \phi} \rangle_{Liouv}
 \cr
 \sim  &  ~ i \mu^{ 2 R -1 } R^2 { \Gamma(- R) \over \Gamma(R) }
\cr
\langle 2pt \rangle = & -  \mu^{ 2 R } R^2 { \Gamma(- R) \over \Gamma(R) }
}}
where the last line is the  answer in the full string theory computation, including the
contribution due to the Euclidean time direction.

We can now consider the computation of the euclidean partition function
for the 0B matrix model in the adjoint representation. In this formula
we will have to add over the two separate phase shifts
\eqn\zerobadj{\eqalign{
Z_W^{0B} =&  \int d\epsilon { \partial_\epsilon (\delta_+(\epsilon ) + \delta_-(\epsilon) )
\over 2 \pi}
e^{ - 2 \pi R \epsilon}
= - \mu^{2 R} \int_{-\infty}^\infty   d\hat \epsilon  \half \hat
\epsilon ({ 1 \over \tanh \pi \hat \epsilon} +1)
   e^{ - 2 \pi R \hat \epsilon}  =
   \cr
   & = -{ 1 \over 4 \sin^2 \pi R } \mu^{ 2 R}
     =
    - { \mu^{ 2 R} \over 4 \pi^2}R^2  \Gamma(-R)^2   \Gamma(R)^2
}}
We see that we get an answer that is essentially the same as the bosonic string
\resinteg , up to the replacement $\mu^R \to \mu^{2 R}$. Again, this is the
proper normalization.

We can also reproduce the result \zerobadj\ by considering the cylinder diagram
between two FZZT branes extended in time. This cylinder diagram  comes purely  from
the exchange of NS-NS closed strings only because these FZZT branes
do not have RR charge in 0B.
 This exchange has a form similar to \cylbos\ so the situation
is very similar to what we encountered in the bosonic case.

\newsec{Non-singlets in the complex matrix model and 0A superstrings}

Two dimensional  0A superstring theory was conjectured to be dual to a $U(N)\times U(M)$
gauged matrix model, where the matrix $M$ is in the $(N , \bar M)$ representation
of the gauge group \chatone .
In this section we consider the model in the presence of non-trivial representations.
So we can choose representations ${\cal R}$, $\tilde {\cal R}$ of the two gauge
groups and write
\eqn\lagrang{
Z = \int {\cal D} m {\cal D} A {\cal D} \tilde A e^{ i  \int Tr[ (D_0 m)^\dagger (D_0 m)  +
V(m^\dagger m) ]}
Tr_{\cal R}  e^{ i \int A } Tr_{\tilde {\cal R} } e^{ i \int \tilde A }
}
In a gauge where we set $A_0=0$ then we simply have a complex matrix model with
the requirement that we keep only the states which transform in representations
${\cal R}^\dagger , {\tilde {\cal R}}^\dagger$ under $U(N) \times U(M)$.
Now we can ask which representations are allowed.
We obviously have that $U(N) \times U(M) \sim U(1) \times SU(N) \times
\tilde U(1) \times \tilde
SU(M)$. Clearly the representations cannot transform under the overall $U(1)$, so
$Q = - \tilde Q$.
In this paper we will consider representations whose charges, including $Q, \tilde Q$,
 remain fixed as $N\to \infty$.
We will also keep $M-N$ finite\foot{ It is also interesting to consider the case
when $Q = - \tilde Q \sim \hat q N$, which correspond to adding a certain RR flux
in the string theory. We consider this case in a separate publication \jmns. }.
In this case, we have that
\eqn\constrq{
\tilde Q = Q_N = - Q =- \tilde Q_M
}
where $Q_N$ and $\tilde Q_M$ are the N-alities of the representations of $SU(N)$ and
$SU(M)$ respectively. If we keep the representations ``fixed'' when $N\to \infty$ we
can think of the N-alities as well defined integers.

Using methods similar to the ones used to derive \hamiltred\ we can derive the
reduced Hamiltonian for a complex matrix acting on wavefunctions that
transform under representations ${\cal R}$, $\tilde {\cal R}$
\eqn\hamiltgen{\eqalign{
H = & \left[\sum_{i=1}^N - \half { \partial^2 \over \partial \rho_i}  - \half \rho_i^2 +
\half { (\Pi_{i}^i)^2 + (N-M)^2 - { 1 \over 4} \over
\rho_i^2}  + \right.
\cr
+ & \left.
2 \sum_{ i <  j \leq N} { (\rho_i^2 + \rho_j^2) (
\Pi_j^{ ~ i} \Pi_i^{~ j} + \tilde \Pi_j^{~ i} \tilde \Pi_i^{~ j} ) +
2 \rho_i \rho_j (\Pi_j^{~ i} \tilde \Pi_i^{~ j} +
\tilde \Pi_j^{~ i} \Pi_i^{~ j} ) \over (\rho_i^2 - \rho_j^2)^2 }
+ \right.
\cr
+ & \left. \sum_{i=1}^N  { 1 \over \rho_i^2}\sum_{j>N} ( \tilde \Pi_j^{~ i} \tilde \Pi_i^{~ j} +
\tilde \Pi_i^{~j} \tilde \Pi_j^{~ i} ) \right] P_0
}}
where $P_0$ is a projector on the states obeying
\eqn\projoa{\eqalign{
 & \Pi_{i}^i + \tilde \Pi_{i}^i =0 ~~~~({\rm no~sum})
 \cr
 & \tilde \Pi_l^{~ k} =0 ~,~~~~~l,k>N
 }}
where $\Pi_{ij}$ are the $U(N)$ generators and $\tilde \Pi_{ij}$ are the $U(M)$
generators. The last line implies that under the decomposition
$SU(N) \times SU(M-N) \times U(1)
\subset SU(M)$ we consider only states that are trivial under $SU(M-N)$. I have
assumed that $M\geq N$.

\subsec{Two simple examples of non-trivial representations}

 Consider $N=M$ and pick a state in  the adjoint of $SU(N)$ and
the trivial representation in $\tilde U(M)$. So $\tilde \Pi_{i}^{\, j} =0$.
Then $\Pi_{ii}=0$ (no sum) and the Hamiltonian becomes
\eqn\hambe{
H = \sum_i \left( - \half \partial_{\rho_i}^2  - \half \rho_i^2 + \half { (- { 1 \over 4})
\over \rho_i^2}\right) + 2 \sum_{i<j} { \rho_i^2 + \rho_j^2 \over (\rho_i^2 - \rho_j^2)^2}
\Pi_i^{\, j} \Pi_j^{\, i}
}
The term involving the one quarter is subleading at large $\mu$, so I neglect it\foot{
It is not difficult to include it \chatone .}.
As in the hermitian matrix model we can now write
the
eigenvalue problem as
\eqn\oaeignp{
E w(\lambda) = 2 \int d\lambda' \rho(\lambda') { \lambda^2 + \lambda'^2 \over
(\lambda^2 - \lambda'^2)^2} (w(\lambda) - w(\lambda'))
}
We then use a formula similar to \weha\ to find a kinetic term and a potential which
turns out to be exactly the same as what we had for the 0B model.
The crucial aspect is that we can write
\eqn\rewr{
 2 { \lambda^2 + {\lambda'}^2 \over (\lambda^2 - \lambda'^2)^2} =
 { 1 \over (\lambda - \lambda')^2 } + { 1 \over (\lambda + \lambda')^2 }
 }
So the problem is basically the same as the problem we had for the
0B case for the so called even functions.
We expect then that the scattering phase is given by $\delta_+$ in \twopos .
Obviously, we would get the same answer if we considered a representation which
is trivial in $SU(N)$ and is the adjoint of $S\tilde U(N)$.

Another interesting representation is
a fundamental of the first and
an anti-fundamental of the second gauge group. Again we can label the states
in the representation by $W_{i}^j$.  Then the constraint on the
diagonal part forces $W$ to be diagonal.
  We then find an eigenvalue equation
of the form
\eqn\eqnsimp{
E w_i = {\half} { 1 \over \lambda_i^2} w_i +  w_i \sum_j ( { 1\over
(\lambda_i-\lambda_j)^2 } +{ 1\over
(\lambda_i+\lambda_j)^2 } )  - 2 \sum_j   ( { 1\over
(\lambda_i-\lambda_j)^2 } - { 1\over
(\lambda_i+\lambda_j)^2 } )w_j
}
where we used that
\eqn\writt{
 { 4 \lambda \lambda' \over (\lambda^2 - \lambda'^2)^2} = { 1
 \over (\lambda - \lambda')^2 } - { 1 \over (\lambda + \lambda')^2}
 }
So we get a problem which is similar to the 0B problem for odd wavefunctions.
The only difference is a potential term which
goes like $1/\lambda^2$.  This term is subleading in the
$1/\mu$ expansion, so we ignore it. In conclusion, we expect that the scattering
phase is given by $\delta_-$ in \twopos.

\subsec{Partition functions and cylinder diagram for FZZT branes}

Using the phase shifts,
we can easily compute the partition functions for the $\pm$  cases. We obtain
\eqn\zapat{\eqalign{
Z^{+} \sim & \int d\epsilon { \partial_\epsilon \delta_+(\epsilon) \over
2 \pi} e^{ - 2 \pi \epsilon}  = - \mu^{2 R}{ \cos 2 \pi R \over 2  \sin^2 2 \pi R}
\cr
Z^{-} \sim & \int d\epsilon { \partial_\epsilon \delta_-(\epsilon) \over
2 \pi} e^{ - 2 \pi \epsilon}  = - \mu^{2 R}{ 1\over 2 \sin^2 2 \pi R}
}}
Of course, $Z^+ + Z^-$ gives us again \zerobadj .

Now let us consider the vortex anti-vortex correlator in 0A theory. Winding modes
can be in the NS-NS or RR sector. The NS-NS correlator was computed in \twons .
 In order to compute the RR correlator  we  consider the derivative
with respect to $\mu$ of the two point function. We then have
to compute a three point function of one NS and two RR vertex operators. We
put the NS operator in the $-1$ picture and the RR operators in the $-1/2$
picture. So we will need the super Liouville correlator in \fukuda \foot{It is the one
called $\tilde C_3$ in \fukuda .}. We obtain
\eqn\tworr{\eqalign{
 \partial_\mu \langle 2pt  \rangle \sim  & -
 R \mu^{ 2 R -1 }  { \Gamma(\half - R)^2 \over \Gamma(\half + R)^2 }
\cr
\langle 2pt \rangle = & -  \mu^{ 2 R }
{ \Gamma( \half - R)^2 \over \Gamma(\half + R)^2 }
}}

We can also compute the cylinder diagrams for FZZT branes in the 0A theory.
In the 0A case the FZZT branes carry RR charge. So the cylinder diagrams are a bit
more interesting than in the 0B case. The NS contribution is similar to that of the
0B theory, and the bosonic theory. The RR contribution depends
on the sign of $\eta $. For example, for $\eta  = -1$ we have
\eqn\oaetamone{\eqalign{
Z_{cyl}^{\eta = - 1} =&  ( 2 \pi R)\sum_{n} \int dE
\left( {  \cos^2 2 \pi E \sigma \over \sinh^2 \pi E}
+ { \cos^2 2 \pi E \sigma \over \cosh^2 \pi E } \right) { 1 \over E^2 + n^2 R^2}
\cr
\left. Z_{cyl}^{\eta = -1}\right|_{n=1} \sim  &
- \left({1 \over \sin^2 \pi R} - { 1 \over \cos^2 \pi R} \right) \mu^{2R} \mu_B^{-4 R}
}}
where in the second line we have extracted
 the contribution from the first pole at $E = i R$.

Similarly for the $\eta  =1$ brane we get
\eqn\oaetapone{\eqalign{
Z_{cyl}^{\eta = 1} =&  (2 \pi R) \sum_{n} \int dE  \left( { \cos^2 2\pi E \sigma \over \sinh^2 \pi E}
- { \sin^2 2 \pi E \sigma \over \cosh^2 \pi E } \right) { 1 \over E^2 + n^2 R^2}
\cr
\left. Z_{cyl}^{\eta = 1}\right|_{n=1} \sim  & - \left( {1 \over \sin^2 \pi R} + {1 \over \cos^2 \pi R}
\right) \mu^{2R} \mu_B^{-4 R}
}}
In the computation of the cylinder diagrams we did not keep track of the overall (positive)
numerical normalization. Up to this constant they agree with \zapat , after
we identify $Z^+ \sim Z^{\eta =-1}_{cyl}|_{n=1}$ and $Z^- \sim Z^{\eta =-1}_{cyl}|_{n=1}$.
The second term in \oaetamone \oaetapone\ comes from the RR exchange and it has a form
that is similar to \tworr .

\ifig\openspectrum{  Spectrum of open strings between ZZ branes and FZZT branes extended
in time in 0A.
In (a) we see the spectrum between ZZ branes and FZZT branes with the same $\eta$.
In this case the ZZ and FZZT branes are charged under the same gauge field.
In (b) we consider an FZZT brane with opposite $\eta$. Here the ZZ and FZZT branes
are charged under a different gauge field. In both cases only the region where the
FZZT branes dissolve are charged. This region is represented here as a dot, but it is
not sharply localized in target space.}
{\epsfxsize 3.0 in\epsfbox{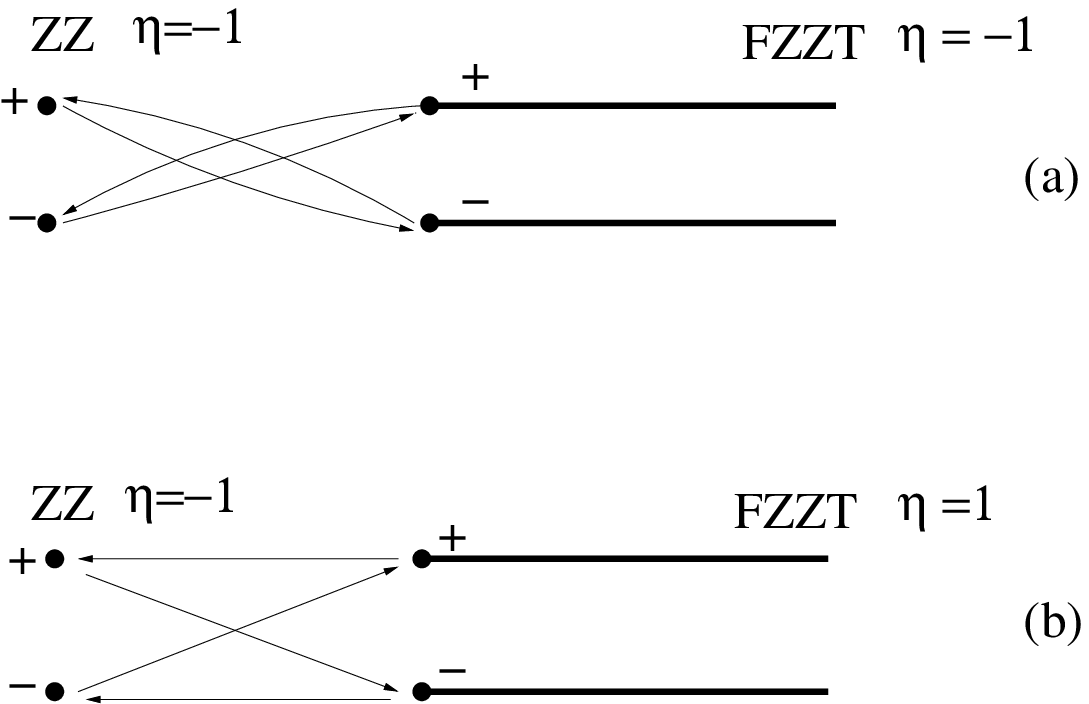}}

What these results are saying is that the open strings that stretch between the
FZZT and ZZ branes should have the following properties. See \openspectrum .
 Let us set $\mu >0$.
We have ZZ branes with two possible charges. Let us consider the open strings
stretching between the ZZ branes and the FZZT branes with the same $\eta$ as the ZZ
branes. In this case we expect that the open strings go between the FZZT brane and
the ZZ brane of a given charge, say plus. The charge they go to is related to the charge
living at the tip of the FZZT brane.
Now let us consider
the open strings between the ZZ and the FZZT brane with the
opposite $\eta$ than the ZZ brane.  We expect that the open strings end on the charge
plus or minus ZZ brane depending on their orientation. Notice that in this case the
ZZ and FZZT branes are charged under different spacetime gauge fields.

Could we see this properties directly from the string theory?. Let us think about
Liouville theory as a rational conformal field theory. Then according to the
Cardy construction the various branes are
labeled by the primary fields. The two ZZ branes are labeled by $ 1, \psi $,
depending on their RR charge. The FZZT branes with the same $\eta$ as the ZZ branes
 will be labeled by
$ e^{ (1 + i \sigma) \phi} $ and $\psi e^{ ( 1 + i \sigma) \phi}$. The open string
spectrum is given by taking the OPE of these fields. We see that this produces
states that have the form $e^{(1 + i \sigma) \phi}$ or $\psi e^{(1+ i \sigma) \phi }$.
For generic $\sigma$
only one of these two states is a superconformal primary. After restricting to superconformal
primary states we see we have the open string spectrum with the properties we sketched
above.
FZZT branes with the opposite $\eta$ are labeled by Ramond states of the form
$R_\pm e^{ ( 1 + i \sigma) \phi}$. In this case the OPE with the states $1, \psi$
will produce states of the form $R_{\pm} e^{ ( 1 + i \sigma) \phi}$ and $R_{\mp}e^{(1 + i \sigma)
\phi}$. Again only half of these states are superconformal primaries annihilated by the
supercharge $G_0$. When we include $X$ the surviving states will depend on the orientation
of the string. This produces a string spectrum which is consistent with the one
we needed on the basis of the matrix model answers.

\newsec{Discussion}

We studied the matrix model in its non-singlet sector. After subtracting a divergence
we saw that the matrix model hamiltonian is unbounded below in this sector.
Nevertheless, there are very natural Lorentzian computations which involve scattering
amplitudes. These amplitudes involve long strings in the target space. These are folded
strings that stretch from the weak coupling region. The tip of the string moves under
the force provided by the string tension. We considered in detail the case of
the adjoint representation (and also bifundamental, for 0A). In this case we have a
single long string. More complicated representations, whose Young diagrams
involve $n$ boxes and $n$ anti-boxes, correspond to states with  $n$ folded long strings.
The string theory results match
the matrix model results,
in the region where we could compute them explicitly.

This picture helps us understand the physical meaning of Euclidean computations
for winding correlators. In the matrix model these are often computed by appealing to
T-duality and then computing momentum correlators in the T-dual free fermion picture.
This helps us understand, for example, the sign of the two point function for winding
correlators.

Probably the picture of non-singlets proposed in this paper will be helpful
for finding a matrix model picture for the Lorentzian black hole. The action
proposed in \kkk\ for the Euclidean black hole does not have an obvious {\it local}
Lorentzian continuation. It is likely that by taking a scaling limit with a large
number of FZZT branes \limscl , we could find a satisfactory lorentzian
picture for the black hole. This point of view would suggest that we should look for
it as an excited long lived state in the non-singlet sector provided
by the FZZT branes in the limit \limscl .

Note that the mode that changes the value of the dilaton at the tip
is not normalizable in $SU(2)_k/U(1)$ for $k<3$ (see section 7.3 of \ref\KarczmarekBW{
  J.~L.~Karczmarek, J.~Maldacena and A.~Strominger,
  ``Black hole non-formation in the matrix model,''
  arXiv:hep-th/0411174.
}). This implies that the thermodynamics of the cigar is qualitatively different than
in the large $k$ situations, which was analyzed in \refs{\kutasovtherm,\berkooz} \foot{
In particular, section 6 of \kkk\ does not seem to
apply to the two dimensional black hole.}. This fact explains that despite appearances
we can actually change the radius of the cigar in two dimensional string theory.
It also explains why there is no sign of the black hole in the singlet sector
thermal partition function \gkthermal .

{\bf Acknowledgements}

I would like to thank N. Seiberg for discussions and for collaboration on
some of the topics discussed.
I also  thank V. Kazakov, I. Kostov and D. Gaiotto for interesting
discussions and for showing me some of their unpublished work on this subject.
I also thank I. Klebanov and D. Kutasov for discussions on these issues.

This work was supported in part by  DOE grant DE-FG02-90ER40542. I would also like
to thank the hospitality of the IHES, where part of this work was done.

\appendix{A}{Derivation of the scattering amplitudes}

\subsec{Bosonic string}

We start by taking the $b\to 1$ limit of the formulas in
\fzzbdy . The two point function for the scattering of two open strings with
two ends on the same D-brane is
\eqn\fzzform{\eqalign{
\langle e^{(1 + i \alpha)  \phi} e^{(1 + i \alpha) \phi} \rangle =&
\mu^{- i \alpha}{ \Gamma( 2i \alpha) G(1-2 i \alpha)G(1 + i \alpha + i s)
G(1 + i \alpha)^2  \over
\Gamma( - 2 i \alpha)
G( 1+ 2 i \alpha) G(1-i \alpha - i s) G(1- i \alpha)^2 } { G(1 + i \alpha - i s ) \over
G(1 - i \alpha + i s ) }
\cr
 \cosh \pi s = & { \mu_B \over \sqrt{\mu} }
 \cr
i g(\alpha) \equiv &  \log { G(1 + i \alpha )  \over G(1-i \alpha)} =   i \half
 \int_{0}^\infty { d t \over t} \left[ {\sin 2 t \alpha \over \sinh^2 t} -  {2 \alpha
 \over t} \right]  = - i \pi \int_0^\alpha d\alpha' { \alpha'  \over
 \tanh  \pi \alpha' }
}}
 In the $b\to 1 $ limit we also need to scale the bulk and boundary
cosmological constants to get a finite answer. In \fzzform\ we present the answer
in terms of the rescaled variables.
Note that the function $g(\alpha)$ has the behavior
\eqn\asympbeh{
g(\alpha) \sim  \mp { \pi \over 2} \alpha^2 ~~~~~~~~~~~~{\rm for}~~~~\alpha \to
\pm \infty
}
up to exponential corrections.
We now take the scaling limit
\eqn\scalinlim{
 s \to \infty ~,~~~~~~~ \alpha \to \infty ~,~~~~~~~~~ \epsilon \equiv \alpha - {1 \over \pi }
 \log 2 \mu_B = \alpha -s  - { 1 \over 2 \pi} \log \mu  = {
 \rm finite}
}
We see that the last factor in the first line in \fzzform\ is finite. There is also
a finite contribution from other terms. This arises because we want to write
every $\alpha $ as $\alpha = s + \hat \epsilon $ and then take the limit.
Using \asympbeh\ we find that the scattering amplitude is
\eqn\scattampl{\eqalign{
e^{i \delta_{FZZ} }  = & e^{i \delta (\epsilon) } e^{i \delta_{def}(\epsilon) }
e^{i \delta_{div}(\epsilon) }
\cr
\delta  = & g(\epsilon + { 1 \over 2\pi } \log \mu) - { \pi \over 2 } (\epsilon
+ { 1 \over 2\pi } \log \mu )^2
\cr
\delta_{def} = & { 1\over \pi}  (\phi_c + \pi \epsilon)^2
\cr
\delta_{div}  = &  -  2 \phi_c (\epsilon + { 1\over \pi } \phi_c)  +
 { 4 \over \pi}   \phi_c (\log 2 \phi_c/\pi  -1) + 4  \epsilon
\log 2 \phi_c/\pi  ~,~~~~~~~ \phi_c \equiv  \log (2 \mu_B)
}}
where the phase $\delta_{def}$ is a phase that appears in the definition of the
scattering phase in \phasedef . The first term in the divergent piece can be
interpreted as coming from the propagation from infinity to
$\phi_c $. The rest of the terms in the divergent amplitude seem
related to the details of the regularization procedure through the FZZT brane. In other
words, they are probably
related to the particular way that the ends of the string interact
with the boundary Liouville potential.
We can forget about the divergent piece.
 Notice that the divergent pieces are independent of $\mu$. This is
a consistency check since all these divergent pieces are related to our particular
regularization of the amplitude using FZZT branes and these branes are dissolving far
away from the region where the Liouville cosmological constant is important.
After defining
\eqn\beg{
\delta( \epsilon) \equiv g(\hat \epsilon) - { \pi \over 2} \hat \epsilon^2
}
we recover \amplitbos .

In the limit when $\epsilon \to -\infty$, then $\delta \to 0$ in agreement with
\classphase .
When $\epsilon \to + \infty $  we get
 \eqn\simplplus{
 e^{i \delta} e^{i \pi \epsilon^2 }  \sim  \mu^{ - i   (\epsilon + {1 \over 4 \pi} \log \mu ) }
 }
 The energy dependence is the precisely the one we get for the scattering of a
 massless ``tachyon" from the Liouville potential. We have included a factor of
 $
 e^{ i \pi \epsilon^2 }$ which comes from the difference between the
 definition of the scattering amplitude as in \phasedef\ versus the more standard
 definition which would lack such factor.  This
 more standard definition is the one   used to study  the
 scattering of massless particles.
 We have chosen the apparently more complicated definition \phasedef\ because with
 this definition the scattering phase depends only on $\epsilon + {1 \over 2\pi } \log \mu$.
 Had we chosen a more standard definition via
 \eqn\morestand{
 \psi \sim  e^{ - i { 1 \over 2 \pi} \phi^2 + i \epsilon \phi} + e^{i\delta_{usual} }
 e^{ i { 1 \over 2 \pi} \phi^2 - i \epsilon \phi}
 }
 then we would find that
 \eqn\relation{
 \delta_{usual} = \delta + \pi \epsilon^2 + \pi
 }
 where $\delta$ is defined in \phasedef . Notice that it is $\epsilon$ that appears here
 and not $\hat \epsilon = \epsilon + { 1 \over 2 \pi } \log \mu  $.
While $\delta$ depends only on $\hat \epsilon$, $\delta_{usual}$ in \relation\ depends
on both $\hat \epsilon $, $\epsilon$, but the dependence on the latter is very simple.

\subsec{ Type 0 superstring}

In order to compute the scattering amplitudes for the superstring we start
with the exact formulas of for the two point function in \fukuda .
Again we just present the $b \to 1 $ limit of the formulas in \fukuda .
Let us consider first the scattering of NS open strings which stretch between
FZZT branes with the same $\eta $.
The result for the scattering of two such open strings depends on $\eta$.
Let us consider, for example, the amplitude for the scattering of open strings
on the $\eta =-1$ FZZT brane, in conventions where the ZZ brane has $\eta  =-1$.
This
gives\foot{ Notice that $2 \phi_{here} = \phi_{Fukuda-Hosomichi}$
in \fukuda .}
\eqn\suplus{\eqalign{
\langle e^{ (1 + i \alpha )\phi} e^{( 1 + i \alpha )\phi} \rangle_{\eta =-1} =&
\mu^{ -  i \alpha } { G( 1-  i \alpha)^2  \Gamma( i \alpha) \over G( 1 +   i \alpha)^2
\Gamma(- i \alpha)}
{ G_{NS}( 1+ i \alpha +  2 i s) \over
 G_{NS}( 1- i \alpha -  2 i s)} {G_{NS}^2( 1 + i \alpha)
 \over G^2_{NS}( 1 - i \alpha) } \times
 \cr
  &  \times {  G_{NS}(1+i \alpha - 2 i s) \over G_{NS}(1 -i \alpha + 2 i s)}
\cr
G_{NS}(x) = &  {G({x \over 2} ) G({x + 1 \over 2} )}
\cr
\cosh \pi s = & { \mu_B \over \sqrt{\mu} }
\cr
i g_{+}(\alpha)  \equiv & \log{ G_{NS}(1 + i \alpha ) \over G_{NS} (1 - i \alpha)} = i
g( {\alpha \over 2}  + {i \over 2} ) + i g ( { \alpha \over 2}  - { i \over 2} ) \cr
 = &
- i 2 \int_0^{\alpha \over 2}  d \alpha' { \alpha' \pi } \tanh \alpha' \pi
}}
where $G(x)$, and $g(\alpha)$ are the same functions as in \fzzform  .
In these formulas $\alpha$ is equal to the energy of the open string.
Using the asymptotic behavior
of $g_{NS}(\alpha)$ as $\alpha \to \pm \infty$ we can take the scaling limit
\eqn\scalimsu{
s \to \infty ~,~~~~~~~\alpha \to \infty ~, ~~~~~~
\epsilon \equiv  \alpha - {2 \over \pi }
 \log 2 \mu_B = (\alpha - 2 s  - { 2 \over 2 \pi} \log \mu ) = {
 \rm finite}
 }
After throwing away divergent pieces and pulling out a piece
  of the required form to define the phase as in \phasedef\ we find
\eqn\finans{\eqalign{
\delta_+ & =  - 2 \int_{-\infty}^{\hat \epsilon/2}
 d \alpha' { \alpha' \pi  } ( \tanh \pi \alpha'  + 1)
\cr
\hat \epsilon &\equiv \epsilon + { 1 \over \pi } \log \mu }}
Note that $\hat \epsilon$ for the superstring is defined differently than
in the bosonic string, due to the factor of two in the value of $\alpha'$.
Note that \finans\  is such that it goes to zero as $\alpha \to 0$, in accordance
with \classphase .

We now turn our attention to the case of an open string that lives on an FZZT
brane with $\eta =1$.
In this case we obtain
\eqn\suplmin{\eqalign{
\langle e^{ (1 + i \alpha )\phi} e^{( 1 + i \alpha )\phi} \rangle_{\eta =1} =&
\mu^{ -  i \alpha } { G( 1-  i \alpha)^2  \Gamma( i \alpha) \over G( 1 +   i \alpha)^2
\Gamma(- i \alpha)}
{ G_{R}( 1+ i \alpha +  2 i s) \over
 G_{R}( 1- i \alpha -  2 i s)} {G_{NS}^2( 1 + i \alpha)
 \over G^2_{NS}( 1 - i \alpha) } \times
 \cr
  &  \times {  G_{R}(1+i \alpha - 2 i s) \over G_{R}(1 -i \alpha + 2 i s)}
\cr
G_{R}(x) = &   G({x \over 2} + \half )^2
\cr
\cosh \pi s = & { \mu_B \over \sqrt{\mu} }
\cr
i g_{-}(\alpha)  \equiv & \log{ G_{R}(1 + i \alpha ) \over G_{R} (1 - i \alpha)} = i
2 g( {\alpha \over 2} )
 =
- i 2 \int_0^{\alpha \over 2}  d \alpha' { { \alpha' \pi } \over  \tanh \alpha' \pi }
}}
where $G_{NS}$ is the same as the one defined above. Notice that, despite the $R$ index,
we are scattering an NS open string. We are just using the notation in \fukuda .
Taking the same limit as in \scalimsu\ we find the phase
\eqn\phaseram{
\delta_- = - 2 \int_{-\infty}^{\hat \epsilon \over 2}  d \epsilon'   \epsilon' \pi \left( { 1
 \over
 \tanh \epsilon' \pi } + 1 \right)
}

\appendix{B}{Computation of matrix model reflection amplitudes}

\subsec{ Derivation of \newform }

We can rewrite \weha\ as
\eqn\equat{\eqalign{
E h(\tau)= & -  {1 \over \pi} \int_0^\infty d\tau'{
\sinh\tau \sinh \tau'  \over (\cosh \tau - \cosh \tau') }
  h(\tau')  + v(\tau) h(\tau)
\cr
v(\tau) =&    { 1 \over \pi}
\int_0^\infty d\tau'   {  \sinh^2 \tau'
 \over (\cosh \tau - \cosh\tau')^2} \cr
1 = & { 1\over \pi} \int_0^\infty d\tau h(\tau)^2
 }}
It is possible to rewrite the kinetic term using
\eqn\kinreq{
{ 1 \over \pi} {  \sinh\tau \sinh \tau'  \over (\cosh \tau - \cosh \tau')^2 } =
 { 1 \over \pi} { 1 \over 4 \sinh^2{\tau - \tau'\over 2} } -
 { 1 \over \pi} { 1 \over 4 \sinh^2{\tau + \tau'\over 2} }
}
Similarly the potential can be written as
\eqn\poteint{
v(\tau)  = { 1 \over \pi} \left( \tau_{max}- 1 -  { \tau \over \tanh \tau }
  \right)
}
Since
the eigenvalue distribution goes to zero at $\tau =0$ it is reasonable to assume
that $h(\tau=0) =0$. Note that $h(\tau)$ is initially defined only for $\tau>0$.
It is reasonable to extend the definition of $h(\tau)$ on the whole real axis by
defining it as an odd function of $\tau$. Namely,
  $h(\tau) = - h(-\tau)$ for $\tau <0$. Then we can
rewrite the kinetic term as an integral in $\tau'$ from minus to plus infinity.
In other words we find that
\eqn\rewri{
-{1 \over \pi} \int_0^\infty   d\tau'{  \sinh\tau \sinh \tau'  \over (\cosh \tau - \cosh \tau') }
 h(\tau')  =
- {1 \over \pi} \int_{-\infty}^\infty d\tau' {
 h(\tau')  \over 4 \sinh^2{\tau - \tau' \over 2} }
}
We are then left with \newform\ \defofe .

\subsec{ Some scattering amplitudes from the matrix model}

Let us discuss solutions of \newform .
Let us first assume that $\hat \epsilon \ll 0$. In this case we can work in Fourier
space and approximate the potential as ${ \tau \over \tanh \tau } \sim \tau = i \partial_k$.
Then we get the equation
\eqn\geteq{
\hat \epsilon h(k) =   \left( { k \over \tanh \pi k } - i { 1 \over \pi } \partial_k \right) h(k)
}
which is solved by
\eqn\solhes{\eqalign{
h(k) = & Exp\left( i \pi \hat \epsilon k -  i \pi \int_0^k dk' { k' \over \tanh \pi k' } \right)
\cr
h(k) ~ & \longrightarrow Exp\left( i \pi \hat \epsilon k \mp {\pi \over 2} k^2 - i \pi/12
\right) ~~~~~~{\rm as} ~~~~~~k\to \pm \infty
}}
Its Fourier transform for large $\tau$ can be done by saddle point approximation
and we get
\eqn\resrealt{
h(\tau) \sim
e^{ - i { 1 \over 2 \pi }(\tau + \pi \hat \epsilon)^2 } -
 e^{ i { 1 \over 2 \pi} (\tau + \pi \hat \epsilon)^2 } ~,~~~~~~~\tau \gg 1
}
where we ignored an overall phase.
We then conclude that the scattering phase is zero.
In general, we are defining the phase through
\eqn\phasedefmm{
h(\tau) \sim
e^{ - i { 1 \over 2 \pi }(\tau + \pi \hat \epsilon)^2 } - e^{ i \delta}
 e^{ i { 1 \over 2 \pi} (\tau + \pi \hat \epsilon)^2 } ~,~~~~~~~\tau \gg 1
}
We can now compute the first correction by expanding the potential term ${\tau
\over \tanh \tau } \sim \tau + 2 \tau e^{ - 2 \tau} $ and
writing $h = h_0( 1 + v_1)$ where $h_0$ is the leading order solution \solhes .
We then get the
equation
\eqn\vonee{
  \partial_k v_1 = - 2     h_0^{-1}(k) \partial_k [ e^{ - 2 i \partial_k } h_0(k)]
 = 2(  -i \pi \hat \epsilon  + i { \pi ( k - 2 i) \over \tanh \pi k} ) e^{ - i \pi \int_k^{k- 2i }
 { k' \over \tanh \pi k' } } e^{ 2 \pi \hat \epsilon}
}
We can evaluate the integral in the exponent of \vonee\ to obtain
\eqn\inrest{\eqalign{
v_1(\infty) - v_1(-\infty) =&  - 2i   e^{ 2 \pi \hat \epsilon} \int_{-\infty}^\infty \left(
  \pi \hat \epsilon  -   \pi { k  -  2 i  \over \tanh \pi k } \right)
{ 1 \over 4 \sinh^2 \pi k }
\cr
= &   i (   \hat \epsilon - {1 \over 2 \pi} ) e^{ 2 \pi \hat \epsilon}
}}
where we integrated by shifting the contour away from the singularity at $k =0$.
We see that we reproduce \limbeh .

Let us now study the opposite regime, namely $\hat \epsilon \gg 0$.
We can approximate the kinetic term as $K(k) = k  $.
Then the equation and the solution becomes
\eqn\leadingap{\eqalign{
0=& ( \hat \epsilon +   i \partial_\tau + { \tau \over \pi \tanh \tau} ) h(\tau)
\cr
h(\tau) = & e^{ i \hat \epsilon \tau + i { 1 \over \pi} \int_0^\tau { \tau' \over
\tanh \tau'} d\tau' }
\cr
h(\tau) \sim & e^{ i \tau \hat \epsilon + i {\tau^2 \over 2 \pi} + i { \pi \over 12 } }  ~,~~~~~~~~\tau \gg 0
\cr
h(\tau) \sim & e^{ i \tau \hat \epsilon - i { \tau^2 \over 2 \pi} + i { \pi \over 12 } }  ~,~~~~~~~~\tau \ll 0
}}
In order to solve the boundary conditions at the origin we still need to impose
that $h(\tau) = h(-\tau)$. This can be done by taking $h \to h(\tau) - h(-\tau)$
with $h$ as defined above. Comparing with \phasedefmm\ we see that the
leading order phase
is $\delta = - \pi \hat \epsilon^2$ in agreement with the leading
order term in \limhe .
Now we can do a first order expansion. We expand the kinetic term as
$E(k) = k + 2 k e^{ - 2 \pi k} $. The exponential
is an operator performing a translation $\tau \to \tau + 2 \pi i$.
As before we write $h = h_0(1 + v_1)$ and we get
\eqn\equforv{
i \partial_\tau v_1 = - 2 i {\partial_\tau h_0(\tau + 2 \pi i)
\over h_0(\tau) } =  2 ( \hat \epsilon + { \tau + 2 \pi i \over \pi \tanh \tau} ) e^{i
\int_\tau^{\tau + 2 \pi i} { \tau' \over \pi \tanh \tau'} d \tau' } e^{ -2 \pi \hat \epsilon}
}
We can evaluate the integral in the exponent of \equforv\ to
 obtain
\eqn\givesans{
v_1(\infty) - v_1(-\infty) = -i e^{ - 2 \pi \hat \epsilon}
\int_{-\infty}^\infty ( \hat \epsilon + { \tau + 2 \pi i \over
\pi \tanh \tau} ) { 1 \over 2 \sinh^2\tau }
= i ( \hat \epsilon  + {1 \over 2 \pi})e^{ - 2 \pi \hat \epsilon }
}
where again we have shifted the contour
to $\tau \to \tau + i \pi $.
We see then that we reproduce the subleading term in \limbeh .

For the two sided matrix model we can also compute the leading terms and the first
correction. The whole computation is very similar to the one we did above, except
for factors of two. When we compute the first subleading correction starting from
high energies $\hat \epsilon \gg 0$, then it is easy to see that the results for
 $\delta_\pm$ differ by just a sign, as we expect from the high energy expansion of
 \twopos\ which is
 \eqn\highenp{\eqalign{
 \delta_- \sim &  - { \pi \over 2} \hat \epsilon^2 +
 ( \hat \epsilon + { 1 \over \pi}) e^{- \pi  \hat\epsilon}
 ~,~~~~~~~~\hat \epsilon \gg 0
\cr
\delta_+ \sim &  - { \pi \over 2} \hat \epsilon^2 -
 ( \hat \epsilon + { 1 \over \pi}) e^{- \pi  \hat\epsilon}
 ~,~~~~~~~~\hat \epsilon \gg 0
 }}
The difference in sign comes from the difference in sign when we expand the kinetic
terms in \kinen , $K_\pm \sim k \mp 2k e^{ -  \pi k} $.

For very low energies $\epsilon \ll 0$ we can do a similar expansion. Clearly for $\delta_-$
we will get a result similar to the one sided model, in agreement with
\leadob . For $\delta_+$ the leading
correction will have the opposite sign which comes, basically, from the fact that
we will have a $(\cosh  \pi k/2)^{-2}$ instead of $(\sinh \pi k/2)^{-2}$
in an expression similar
to \inrest . Going through all the
details one indeed finds the opposite sign, in agreement
with \leadob .

In conclusion, we have checked explicitly that the leading
and first subleading terms computed
in the matrix model agree with the string theory answers.
This holds both for the bosonic string and the superstring cases.
It is interesting that the same functions are appearing in the potential, the
kinetic terms and the phase. This suggests that there must be a simpler way to think
about problem this that would allow us to get the answer more quickly.

\appendix{C}{ The FZZ conjecture}

We start from the  $N=2$ version of the FZZ conjecture, which was considered in
\hk , see also \ref\TongIK{
  D.~Tong,
  ``Mirror mirror on the wall: On two-dimensional black holes and Liouville
  theory,''
  JHEP {\bf 0304}, 031 (2003)
  [arXiv:hep-th/0303151].
}. This version of the conjecture is defined for any value of $k$, regardless
of whether the perturbation away  from the linear dilaton region is or is not normalizable.

We now observe that the  cigar theory with worldsheet supersymmetry
is the same as a bosonic
cigar plus two free fermions. So there should be some free fields also on
the ${\cal N}=2$ sine Liouville side. It is not difficult to identify the free
field, it is just the $U(1)$ current of the ${\cal N}=2$ superconformal
algebra. After quotienting by this current we get the bosonic ${\cal N}=2 $
sine Liouville. So we conclude that the bosonic version of the
SineLiouville/cigar duality derives from the ${\cal N} =2$ version studied in \hk .

Let us see this a bit more explicitly\foot{ In this appendix set $\sqrt{\alpha'} =1$.}.
Let us denote by  $k_s$ the level of the ${\cal N}=2$ cigar theory.
Let us also define  $k_b =
k_s +2$, which is the level of the associated bosonic cigar theory.
Then the $N=2$ Liouville theory   has the interaction
terms
\eqn\interliouv{
\mu e^{i \sqrt{2} \sigma}
e^{ \sqrt{k_s} \phi + i\sqrt{k_s}  (\varphi_L - \varphi_R)}
}
where $\sigma$ accounts for  the bosonized fermions.
We now consider the expression for the ${\cal N}=2$ U(1) current and
we define a new scalar field $x$, which is orthogonal to it
\eqn\defi{\eqalign{
j \sim & \sqrt{2}( \partial \sigma_L - { \sqrt{2} \over \sqrt{k_s}} \partial \varphi_L)
\cr
  \sqrt{k_b} x_L \equiv & \sqrt{2} \sigma_L + \sqrt{k_s} \varphi_L
}}
and a similar expression for right movers.
Using \defi\ we can rewrite \interliouv\ as
\eqn\interbosl{
 \mu e^{\sqrt{k_b -2} \phi + i \sqrt{k_b} (x_L - x_R)  }
 }
 which is precisely the interaction of the bosonic sine-Liouville theory that
 was conjectured in \fzzdual\ (see \kkk ) to be dual to the cigar theory at level $k_b$.

Using more detailed formulas in \ari\ it is possible to show that the quotients we
have done are consistent with the periodicity conditions for the fields in both
theories. In other words we have
\eqn\gaugedmod{\eqalign{
  \left[{ SU(2)_{k_s}/U(1) } \right]_{susy}/U(1)_R =&  SU(2)_{k_b}/U(1)
  \cr
\left( { \cal N } =2 ~~{\rm SineLiouville} \right)_{k_s}/U(1)_R = &
\left( {\rm Bosonic ~SineLiouville } \right)_{k_b}
}}
So the bosonic version of FZZ follows from \hk\ by GKO quotienting  by the R-symmetry.

\appendix{D}{Computations of winding correlators via T-duality}

Let us consider the problem of computing matrix model correlators in the singlet sector.
In the region of large $\lambda$ the matrix model fermion reduces to a relativistic
fermion. Since we are at finite temperature this relativistic fermion lives on the
circle. If we bosonize this fermion we have a scalar field which is the tachyon. This
is a free field in the asymptotic region. We can think of this problem as a
conformal field theory, given by the tachyon  field or the relativistic fermions, with a
boundary. The boundary encodes the small $\lambda$ region and the reflection amplitudes
for the fermions. In the fermion basis the boundary state is very simple, we simply have
a linear transformation between the left and right moving fermion by a phase which is
the scattering phase in the upside down harmonic oscillator potential.

The correlators that we want to compute are most simply expressed by expanding
the tachyon field in momentum modes along the Euclidean time circle. In the asymptotic
region these can be expressed in terms of relativistic fermions.
Again, it is convenient to expand the fermions along the Euclidean time circle. These
will have half integer moding since the fermions are anti-periodic on the Euclidean time
circle.

For example, if we are interested in the first momentum mode of the tachyon we
write it as
\eqn\modet{
\alpha_{-1}|0\rangle  = \tilde \psi_{-1/2} \psi_{-1/2} |0 \rangle
}
where $|0 \rangle $ is the vacuum on the cylinder. We have similar formulas for the
incoming and outgoing tachyon modes corresponding to incoming or outgoing fermions.
In the cylinder incoming or outgoing fields are related to holomorphic or antiholomorphic
fields and
they correspond to modes of the tachyon with positive or negative momentum
on the circle \kos .
So the computation boils down to an expression of the form
\eqn\mdofs{
\langle B | \alpha^{in}_{-1} \alpha^{out}_{-1} |0\rangle
}
The boundary state $|B\rangle$ encodes the reflection amplitudes for the fermions.
These can be expressed through the identities
\eqn\equationsbdy{\eqalign{
 &  \langle B| \left( \psi_{in,s} - i {\cal R}^{-1}(\mu + i s/R) \psi_{out,-s} \right)  =0
~,~~~~~~~~~~~~~s \in Z + \half \cr
& \langle B | \left( {\tilde \psi}_{in,s} - i {\cal R}(\mu - i s/R)
 {\tilde \psi}_{out,-s} \right)  =0
}}
with the standard anticommutations relations
\eqn\commueu{
\{ \psi_{s}, {\tilde \psi}_{s'} \} = \delta_{s+s'}
}
for both in and out fields.
The bounce factors are
\eqn\bounce{
{\cal R}(E) =   \sqrt{
\Gamma( \half - i E) \over \Gamma(\half + i E)}
}
for the bosonic string.

Using these formulas it is easy to compute the correlator in \mdofs . We first
express it in terms of fermions through \modet\ and then use \equationsbdy .
The net result is
\eqn\netres{
\langle B | \alpha^{in}_{-1} \alpha^{out}_{-1} |0\rangle =
\langle B |   |0\rangle {\cal R}(\mu + i { 1 \over 2 R} )
 {\cal R}^{-1}(\mu - i { 1 \over 2 R} )
}
The overlap $ \langle B | 0 \rangle $ is defined to be the thermal partition
function, as computed in \gkthermal .

In summary, \netres\ corresponds to the correlation function for the insertion of
a tachyon with momentum one and a tachyon with momentum minus one along the thermal circle.
After doing the T-duality, which corresponds to sending $R \to 1/R $ and $\mu \to \mu R$
we obtain the two point correlation function for a tachyon with winding number one
and a tachyon with winding number minus one. We identify this with the partition
function of the matrix model in the adjoint representations. In this fashion we
have derived formula \correl .

It is clear that we can use this method to compute more complicated correlators
of winding operators and hopefully it will be clear to the reader how to proceed.

In this way we could imagine computing the partition function in any representation.
All we would need to do is to express the representation in terms of the
winding modes. This can be done using the ideas in \douglascft . We will not
go into the details because there is a subtlety that gets in the way.
If the decomposition involves modes with different winding number, then the
leg factors will be different for these different operators. Since we do not
understand the leg factors we cannot confidently assign a representation
to a particular configuration of winding correlators.
In \ref\maldac{ J. Maldacena, Talk at strings 2004} this subtlety was ignored.

\listrefs

\bye